\documentstyle[a4,epsf,amstex]{article}

\newcommand{\Sv}{{\bf S}}
\newcommand{\Si}{{\boldsymbol \Sigma}}
\newcommand{\pii}{{\boldsymbol \pi}}
\newcommand{\R}{{\bf R}}
\newcommand{\A}{{\cal A}}

\begin{document}
\thispagestyle{empty}
\parskip=12pt
\raggedbottom

\def\mytoday#1{{ } \ifcase\month \or
 January\or February\or March\or April\or May\or June\or
 July\or August\or September\or October\or November\or December\fi
 \space \number\year}

\noindent
\hspace*{9cm} BUTP--98/11\\
\vspace*{1cm}
\begin{center}

{\LARGE The theoretical background and properties of perfect actions}
\footnote{Work supported in part by Schweizerischer Nationalfonds and
by Iberdrola, Ciencia y Tecnologia, Espa\~na.}

\vspace{1cm}
Peter Hasenfratz 
\\
Institute for Theoretical Physics \\
University of Bern \\
Sidlerstrasse 5, CH-3012 Bern, Switzerland

\vspace{0.5cm}
\mytoday \time \\ \vspace*{0.5cm}

\nopagebreak[4]

\begin{abstract}
This lecture note starts with a pedagogical introduction to the theoretical 
background and properties of perfect actions, gives some details on topology
and instanton solutions and ends with a discussion on the recent
developments concerning chiral symmetry.
\end{abstract}

\end{center}
\eject

\section{Introduction}
This lecture note is a summary of the authors contribution at different
schools in 1997: 'NATO ASI - Confinement, Duality and Non-perturbative
Aspects of QCD' in Cambridge, 'Advanced School on Non-Perturbative Quantum
Field Physics' at Pe\~niscola and 'Yukawa International Seminar on
Non-Perturbative QCD' in Kyoto. It is extended by discussing in some detail
recent developments on chiral symmetry.

This lecture note is intended to be largely
self-contained. The general aspects of lattice regularization and Wilson's
renormalization group \cite{wil}
and in particular those points which are closely related to our
subject will be explicitly introduced and discussed. The reader might even
consider our topic as a pretext to learn about these
powerful theoretical and practical tools in quantum field theories. I hope,
however, that the unexpected and amazing properties of the perfect actions
will also catch the readers' fantasy.

The very definition of a quantum field theory requires a regularization. As it
is well known, all the regularizations break some of the symmetries of the
underlying classical theory. They introduce a new scale (the cut-off) which is
reflected in the final predictions even after the regularization is
removed: naive dimensional analysis breaks down, anomalous dimensions are
created, dynamical mass generation might occur, etc. On the other hand, beyond
these exciting physical effects, the new scale creates unwanted 
cut-off artifacts everywhere which have to be eliminated in a theoretically 
and practically difficult limiting process. On the tree level 
(in the classical theory) the new introduced scale creates artifacts only, 
nothing else. If the classical theory
is scale invariant (like the Yang-Mills theory or massless QCD) the cut-off
artifacts destroy the most interesting and relevant classical properties of
the model like the existence of scale invariant classical solutions or
fermionic zero modes and index theorem.

Regularizations often break other symmetries also. The lattice, the only known
non-perturbative regularization, breaks Euclidean rotation symmetry (Lorentz
symmetry)  which is less of a headache, and also chiral symmetry which creates
a difficult tuning and renormalization problem. Chiral symmetry awakes deep
theoretical issues (anomalies, no-go theorems) culminating in problems
concerning chiral gauge theories.

Perfect actions \cite{hn} offer a bold approach to these problems.
We shall call a lattice regularized local action \textit{classically perfect} if
its classical predictions (independently whether the lattice is fine or
coarse, whether the resolution is good or bad) agree with those of the
continuum. The \textit{quantum perfect action} does the same for all the physical
questions in the quantum theory. That such actions exist might seem to be
surprising. Their existence is closely related to renormalization group
theory. As we shall see, they have beautiful properties which
one would not expect a lattice action can have.

The classically perfect action has scale invariant instanton solutions, has no
topological artifacts, satisfies the index theorem concerning the fermionic
zero modes on the lattice, preserves all the important physical
consequences of chiral symmetry in the quantum theory and is expected to
reduce the cut-off effects significantly even in quantum simulations.

In this note we shall mainly discuss the theoretical properties of the
classically perfect actions. These actions are the fixed points of
renormalization group transformations in asymptotically free theories and are
determined by classical field theory equations. They are not abstract
theoretical constructions only. Actually, most of the effort during the last
four years was directed towards constructing and parameterizing them
explicitly and testing their performance in 
numerical simulations \cite{has}. We shall
not discuss these developments here at all beyond the remark that the
fixed point action gives very good results in quantum simulations
including questions related to topology. The main practical question is
whether one can find a parameterization which approximates the fixed point
action sufficiently well without making the simulation prohibitively expensive.

\section{Lattice regularization and the continuum limit}

In this Section we summarize some of the basic notions concerning lattice 
regularization, continuum limit and the relation between Quantum Field Theories
(QFTs) and critical
phenomena in classical statistical mechanics. Readers, who are familiar with
these basics are advised to move directly to Section 2.4 on locality. 

\subsection{QFTs defined by path integrals in Euclidean space}

In these lectures we shall consider QFTs in their Euclidean formulation
and use path integrals. Using path integrals to describe quantum theories
not only provides an appealing physical picture for the quantum evolution
as a sum over classical paths (in quantum mechanics) or sum over classical 
configurations (in QFT),
but gives a powerful framework for analytic manipulations and opens the way
for numerical calculations also. In addition, it helps to understand the
deep relation between QFTs and critical phenomena in classical statistical
mechanics \cite{lenin,text1}.

The formal relation between QFTs and classical statistical physics is
easy to see. Consider an $n$-point function in a scalar QFT in Minkowski space
\begin{equation}
\langle 0|T\bigl(\hat{\phi}(x_1)\hat{\phi}(x_2) \ldots \hat{\phi}(x_n)\bigr)|0\rangle\, ,
\label{spr3}
\end{equation}
where $\hat \phi(x)$ is the field operator. In the path integral 
language this expression is given by
\begin{equation}
\frac{1}{Z}\int D\phi \, \phi(x_1)\phi(x_2)\ldots \phi(x_n)\exp\bigl(i\int dt d{\bf x}~
{\cal L}(\phi,\partial_{\mu}\phi)\bigr),
\end{equation}
where $Z=\int D\phi \exp(i\int dt d{\bf x}\,{\cal L})$ provides for the correct
normalization, $x=(t,{\bf x})=(t,x_1,\ldots,x_{d_s})$, and
the Lagrangian is defined as
\begin{equation}
{\cal L}_{Mink}=\frac{1}{2}\partial_0\phi \partial_0\phi - \sum_{i=1}^{d_s}
\frac{1}{2}\partial_i\phi \partial_i\phi - V(\phi) \; . 
\label{lagr}
\end{equation}
The relation we are looking for becomes clear if we go to Euclidean space
by rotating time to imaginary time
\begin{equation}
t=x_0 \rightarrow -ix_d
\label{wick1}
\end{equation}
It can be shown \cite{schw} 
that in every order of perturbation theory the analytic
continuation in  eq.~(\ref{wick1}) (Wick rotation) is possible due to the
specific pole structure of the Feynman propagators. It is an assumption
that this remains true beyond perturbation theory. Using
\begin{equation}
i \int dt d{\bf x} \rightarrow \int d^d x ,
\end{equation}
and
\begin{equation} 
{\cal L}_{Mink} \rightarrow - \bigl( \, \sum_{\mu=1}^{d}
\frac{1}{2}\partial_\mu \phi \partial_\mu \phi + V(\phi)\bigr)=-{\cal L}_{Eucl},
\label{wick2}
\end{equation}
we get
\begin{equation}
Z=\int D\phi \exp\bigl(-\int d^d x \, {\cal L}_{Eucl}(\phi,\partial \phi)\bigr),
\end{equation}
\begin{equation}        
\langle \phi(x_1) \ldots \phi(x_n)\rangle=
\frac{1}{Z}\int D\phi \, \phi(x_1)\ldots \phi(x_n)
\exp(- \int d^d x \,{\cal L}_{Eucl}).
\label{qftstat}
\end{equation}
These equations can be interpreted as the partition function and the
correlation function of a system in classical statistical mechanics. The
quantum theory of fields in $d_s$ space dimensions is transformed into 
classical statistical mechanics of fields in $d=d_s+1$ Euclidean dimensions.
The Euclidean action plays the role of $E/k_BT$, where $E$ is the classical
energy of the $d$ dimensional configuration.

Actually, the relation between QFTs and classical statistical systems
is deeper and more specific than the observation above. We shall return
to this question in Sect.2.3

\subsection{Lattice regularization}

Some of the mathematical operations which enter the definition of a QFT
require a careful limiting procedure. In field theory, the variables are
associated with space-time points. As can be seen from the form of the
Lagrangians of different classical FT's (see, for example eq.~(\ref{lagr})),
these variables have some kind of self-interaction, whereas the elementary 
interaction between different degrees of freedom is over infinitesimal
distances as expressed by derivatives. Already in the classical theory, the
definition of a derivative requires the temporary introduction of a finite
increment (of the argument of the function considered) which disappears
at the end by some limiting procedure
\begin{equation}
f'(x)=\lim_{a \to 0} \frac{f(x+a)-f(x)}{a}\,.
\label{der}
\end{equation}
Defining the derivative, which has the important role of providing 
communication between different field variables, is a much less trivial
problem in QFTs. Like in eq.~(\ref{der}), the very definition of a QFT
requires the temporary introduction of a defining framework called
regularization, which disappears from the theory by a limiting process.
The way to introduce and remove the regularization is a highly
non-trivial problem. 

We shall use lattice regularization in these lectures. A hypercubic lattice
is introduced in the $d=d_s+1$ dimensional Euclidean space. Scalar and fermion
field variables live on the lattice points, vector fields on the connecting
links. Derivatives are replaced by some kind of finite difference operation.

It will be convenient to work with dimensionless quantities. In the 
following, all the fields, momenta, masses, couplings and other possible
parameters will be defined dimensionless by absorbing appropriate powers
of the lattice unit $a$ in their definition. For any quantity $Q_{df}$ with
dimension $s$ (measured in mass units) we define
\begin{equation}
Q_{df} \rightarrow Q=a^{s} Q_{df}\, .
\label{dim}
\end{equation}
For a scalar field $\phi(x)$ we write
\begin{equation}
\phi(x) \rightarrow \phi_n \, ,
\label{field}
\end{equation}
where $n$ is a $d$-dimensional vector of integers. The simple choice for the
finite difference operation
\begin{equation}
\partial_\mu \phi(x) \rightarrow \nabla_\mu \phi_n = \phi_{n+\hat{\mu}}-\phi_n
\label{nnder}
\end{equation}
leads to the following lattice regularized form of the classical Euclidean
Lagrangian
\begin{equation}
{\cal A} = \int d^dx \, {\cal L} \rightarrow \sum_{n} \sum_{\mu}
\frac{1}{2}\nabla_\mu \phi_n \nabla_\mu \phi_n + \sum_{n} V(\phi_n)\,.
\label{latlagr}
\end{equation}

Having a finite lattice unit $a$ the lattice defines a short distance
regularization. Going over to momentum space, in the Fourier integral
$\exp(ipn)$ enters which is periodic under $p_{\mu}\rightarrow p_{\mu}+2\pi$.
This constrains the momentum to the Brillouin zone 
$- \pi \geq p_{\mu} < \pi$. 
Therefore, the lattice provides a momentum cut-off also.

The lattice regularization has several advantages over other conventional 
regularizations used in perturbation theory. In general, in problems
which require numerical analysis (e.g. integrals, 
differential equations,...),
the standard procedure is to introduce meshes. Also the path integral can 
be defined in a natural way by introducing meshes. The path integral 
regularized this way becomes a set of integrals and is ready for numerical
procedures in case of non-perturbative problems. Actually, lattice
regularization is the only known non-perturbative regularization.  It is a 
very natural regularization also for theories with gauge invariance or with 
constrained variables (like the non-linear $\sigma$-model).

Undoubtedly, lattice regularization has some disadvantages also. It breaks
certain symmetries, as every regularization does. But the lattice has 
special problems (although not unsurmountable) with chiral symmetry and
tough problems with chiral gauge theories. The discretization of space-time 
breaks Lorentz symmetry (O(d) symmetry in Euclidean space), but,
as we shall see later (Sect.~8.1),
the remaining cubic symmetry is sufficient to have
it restored automatically as the regularization is removed (at least in
asymptotically free theories). 
The lack of
Poincar\'e symmetry creates problems, however, in theories with supersymmetry.

Although the lattice is used, in general, to investigate non-perturbative
problems, for certain questions (renormalization of operators, finding
improved actions or comparing lattice results with those obtained in 
other regularizations) it is unavoidable to use lattice perturbation theory.
Unfortunately, perturbation theory on the lattice is technically
somewhat cumbersome.

Some aspects of lattice regularization will be discussed in Sect.~5. For
further reading we refer to textbooks \cite{mont}. 

\subsection{The continuum limit of QFTs versus critical 
phenomena\\ in statistical
mechanics}

Assume, we can solve our lattice regularized QFT at some set of (bare)
couplings and masses in the Lagrangian. The predicted spectrum will contain 
dimensionless numbers, since all our quantities were made 
dimensionless  by absorbing appropriate powers of the lattice 
unit $a$ (Sect. 2.2). The mass of an excitation is given by a number $M$,
the corresponding dimensionfull mass $M_{df}, [M_{df}]=g$ and 
correlation length $\xi_{df}, [\xi_{df}]=cm$ are given by
\begin{equation}
M_{df} = M \frac{1}{a}\, \qquad \xi_{df}=\frac{1}{M_{df}}=
\frac{1}{M}a=\xi a \, .       
\label{mxi}
\end{equation}
The correlation length defined above is the length scale over which the 
massive particle can propagate with a significant amplitude. For a generic 
value of the couplings and mass parameters
in the Lagrangian the predicted $M$ will 
be an $O(1)$ number, the mass of the particle is of the order of the cut-off
$1/a$, the correlation length is $O(a)$. This corresponds to a very poor 
resolution, the situation is far from the continuum limit. It requires a careful
tuning of a certain number of parameters in the Lagrangian to make the
predicted $M$'s in the low lying spectrum to become very small numbers
which correspond to physical length scales much larger than the lattice
unit $a$. In this limit the lattice becomes very fine and the presence 
of the regularization will
not distort the physical results (like mass ratios) anymore.This is the 
process of removing the lattice regularization which leads to the
continuum limit. Theories in which this can be achieved by tuning a
finite number of parameters (and by fixing the normalization of fields
appropriately) are called renormalizable. This criterion is identical
to that introduced in perturbation theory.

In the continuum limit the correlation length of physical excitations
is much larger than the lattice unit $a$. But how large is it in Fermi? This
can not be predicted, the absolute scale should come from observations. If the
theory is expected to describe the physics of hadrons, for example, then
the lowest lying particle with spin=1/2, electric charge=1, 
baryon number=1, strangeness=0,... should be identified with the proton
having a mass of 940 MeV. This fixes the physical length scale
and relative to that unit the lattice constant goes to zero in the 
continuum limit.

In Sect. 2.2 we observed a formal relation between QFTs and classical
statistical systems. We see now that the continuum limit of a QFT corresponds
to a statistical system with a correlation length which is much larger than
the lattice unit $a$. This is the case for critical statistical systems.
In critical solid state problems the lattice unit is fixed 
(typically $O$(\AA)) and the
correlation length  becomes macroscopical. This is a matter of choosing
the absolute scale, however, and it does not reduce the extent of the
deep analogy between the two subjects.

\subsection{Locality}

A recurrent issue in our discussion will be the locality of the action density.
The laws of classical physics (from Newton through 
Maxwell to Einstein) are expressed in terms of differential
equations (rather than, say, integro-differential equations). They correspond
to actions, where the interaction between variables in different space-time
points extends over infinitesimal distances as expressed by derivatives.
This continues to be true in quantum systems also. No symmetry principles
would prevent us to add a term to the Lagrangian in eq.~(\ref{lagr}) like
\begin{equation}
{\cal L}(x)=\ldots + \int d^dy~f(y)\phi(x+\frac{y}{2})\phi(x-\frac{y}{2})\, ,
\label{lagrnl}
\end{equation}
where $f(y)$ is an even function of $y$ and has an extension $\lambda$
(for example $f(y) \sim \exp(-y^2/2\lambda^2)$), where $\lambda$ is finite
when measured in Fermi, or it might decay only as a 
power in $y^2$. Nevertheless, such models found their place neither 
theoretically, nor experimentally. 

In a QFT with non-local interactions renormalizability (in the sense
discussed in Sect. 2.3) will be lost. In addition, a vital and beautiful
property of local QFTs (shared by critical statistical systems), called
universality will be lost also. Universality means that the physical
predictions become independent of the microscopic details of the Lagrangian,
i.e. they are not sensitive to the detailed form of the interaction at the
cut-off scale. Since our knowledge on interactions much above the scales
of present experiments is very limited, universality is vital to
preserve the predictive power of a QFT.

On the lattice we shall call an action density local, if it has an extension
of $O(a)$. Typically, it is an exponential function of the distance
$|n-n'|$ between the variables at $n$ and $n'$: $\sim \exp (-\gamma |n-n'|)$,
$\gamma=O(1)$. In the continuum limit $(a \rightarrow 0)$, the extension of
the action density measured in physical units goes to zero. On the other hand,
an action density which has a finite width in physical units is non-local. No
such non-local actions will be considered here in the following.

A Lagrangian which has nearest-neighbour interactions only or interactions
which are identically zero beyond a few lattice units is certainly local, but
no physical principle requires to have this extreme case. If this were
necessary to assure the above mentioned nice properties of a QFT (or of a
critical statistical system), 
then no experiment on ferromagnets close to the Curie
temperature would observe universality. Really, the elementary interaction
between the magnetic moments in the crystal decays rapidly, but certainly
does not become identically zero beyond a few lattice spacings.

In the following we shall call an action where the couplings become
identically zero beyond a certain $O(a)$ range ultralocal. We shall also
use the expression 'short ranged' for local actions where the exponential
decay goes with a large $\gamma$.

\section{Renormalization group}

A QFT is defined over a large span of scales from low (relative to the
cut-off) physical scales up to the cut-off which goes to infinity in the
continuum limit. Although field variables associated with very high scales do
influence the physical predictions through a complicated cascade process, no
physical question involves them directly. Their presence and indirect
influence makes it difficult to establish an intuitive connection between the
form of the interaction and the final expected predictions. The presence of a
large number of degrees of freedom makes the problem technically difficult
also. It is, therefore a natural idea to integrate them out in the path
integral. This process, which reduces the number of degrees of freedom, taking
into account their effect on the remaining variables exactly, is called a
renormalization group transformation \cite{wil,others} (RGT).

In this Section we introduce some of the basic notions related to RG theory
and set the notations. Since these lectures are mainly concerned with
asymptotically free (AF) QFTs, our discussion will be biased by this goal.

\subsection{Renormalization group transformation}

For simplicity consider a scalar field theory regularized on the lattice 
in $d$ Euclidean dimensions as
discussed in Section~2.2. The RGT averages out the short distance
fluctuations, i.e. the fluctuations over distances $O(a)$. The lattice is
indexed by $n$, the field associated with the point $n$ is denoted by
$\phi_n$. We introduce a blocked lattice with lattice unit $a'=2a$ whose
points are labeled by $n_B$ and the associated block field is denoted by
$\chi_{n_B}$. The block variable $\chi_{n_B}$ is an average of the
original $\phi$ fields in the neighbourhood of the point $n_B$:
\begin{equation}
\chi_{n_B}= b \sum_{n} \omega(2n_B-n)\phi_n \, ,
\label{avr}
\end{equation}
where $\omega$ specifies the weights of the averaging whose normalization is
chosen as
\begin{equation}
\sum_{n} \omega(2n_B-n) = 1 \, .
\label{omnorm}
\end{equation}
The significance of the scale factor $b$ in eq.~(\ref{avr}) will become clear
later. A trivial choice for $\omega$ might be to take $\omega(2n_B-n)=2^{-d}$
if $n$ is in the hypercube whose center is indexed by $n_B$ and zero
otherwise, fig.~1.

\setcounter{figure}{0}
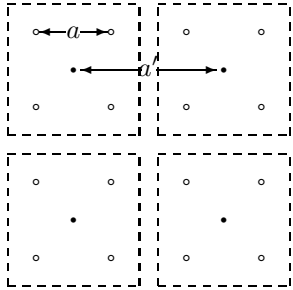
\begin{figure}[htb]
\setlength{\unitlength}{0.5cm}
\begin{center}
\begin{picture}(10,8)
\multiput(2,1)(2,0){4}{\circle{0.15}}
\multiput(2,3)(2,0){4}{\circle{0.15}}
\multiput(2,5)(2,0){4}{\circle{0.15}}
\multiput(2,7)(2,0){4}{\circle{0.15}}
\multiput(1.25,4.25)(4,0){2}{\dashbox{0.2}(3.5,3.5){}}
\multiput(1.25,0.25)(4,0){2}{\dashbox{0.2}(3.5,3.5){}}
\multiput(3,2)(4,0){2}{\circle*{0.15}}
\multiput(3,6)(4,0){2}{\circle*{0.15}}
\put(2.8,7){\vector(-1,0){0.7}}
\put(3.2,7){\vector(1,0){0.7}}
\put(3,7){\makebox(0,0){$a$}}
\put(4.8,6){\vector(-1,0){1.6}}
\put(5.2,6){\vector(1,0){1.6}}
\put(5,6.1){\makebox(0,0){$a'$}}
\end{picture}
\end{center}
\caption{A simple example for a block transformation in d=2.}
\end{figure}

We integrate out now the original $\phi_n$ variables keeping the block
averages fixed:
\begin{equation}
\exp \bigl(-{\cal A}'(\chi)\bigr) = \prod_{n}\int d\phi_n \prod_{n_B}
\delta(\,\chi_{n_B} - b\sum_{n} \omega(2n_B-n)\phi_n\,)\,\exp \bigl(-{\cal A}(\phi)\bigr)\,.
\label{rgeq}
\end{equation}
The new action ${\cal A}'(\chi)$ describes the interaction between the block
variables. The partition function remains unchanged
\begin{equation}
Z'= \prod_{n_B}\int d\chi_{n_B}\exp \bigl(-{\cal A}'(\chi)\bigr)
=\prod_{n}\int d\phi_n \exp \bigl(-{\cal A}(\phi)\bigr) = Z \, .
\label{zpez}
\end{equation}
The long-distance behaviour\footnote{`Long-distance' is meant here
and elsewhere in this work as 'many lattice units', and it might mean short,
or long physical distances.} 
of Green's functions, and so the spectrum and
other low energy properties of the system is expected to remain unchanged as
well. On the other hand, the lattice unit is increased by a factor of 2:
$a'=2a$. Since the dimensions are carried by the lattice unit (Sect.~2.2), the
dimensionless
correlation length, which is measured in the actual lattice distance, is
reduced by a factor of 2:
\begin{equation}
\xi \rightarrow \frac{\xi}{2} \, .
\label{xired}
\end{equation}
By eliminating the fluctuations on the shortest scales we reduced the number
of variables by a factor of $2^d$, while taking their effect into account by
changing the action ${\cal A} \rightarrow {\cal A}'$ appropriately. Although the lattice
unit is increased by a factor of 2, and so the resolution became worse, the
long-distance behaviour remained unchanged. In particular, no new cut-off effects are
generated in the predictions for physical quantities. Iterating this RGT, the
goal formulated at the beginning of Sect.~3 is achieved.

We close this subsection by generalizing the RGT of eq.~(\ref{rgeq}) 
slightly \cite{bell}
(a step which will be very useful later):
\begin{multline}
\exp \bigl(-{\cal A}'(\chi)\bigr)=  \\
{\cal N}(\kappa)\prod_{n}\int d\phi_n
\exp \bigl[-{\cal A}(\phi)-\kappa \sum_{n_B}\,\bigl( \chi_{n_B}-b\,\sum_{n}
\omega(2n_B-n) \phi_n \bigr)^2 \bigr], 
\label{rgeqg}
\end{multline}
where $\kappa$ is a free parameter. For $\kappa \rightarrow \infty$
eq.~(\ref{rgeqg}) goes over to eq.~(\ref{rgeq}), for finite $\kappa$ the block
average is allowed to fluctuate around $\chi$ (slightly, if $\kappa$ is not
too small). Therefore, the parameter $\kappa$ specifies the stiffness of the
RG averaging. ${\cal N}$ is a normalization factor
(which is field independent, trivial in this case)  introduced to keep
eq.~(\ref{zpez}) valid.

\subsection{Constraints on the block transformation 
from \\
symmetries}

It is largely arbitrary how the block averages in a RGT are constructed, but
the procedure should conform with the intuitive goal of a RGT: it should lead
to the elimination of the short distance fluctuations. Identifying, for
example, the block variable with one of the original fine variables in the
block (`decimation') is not really an averaging, and although it is a legal
transformation of the path integral, it will not lead to useful results in
general.

Beyond this general requirement, symmetries put further constraints on the
form of the block transformation. Denote the fields on the fine and coarse
lattice by $\phi$ and $\chi$, respectively and write the RGT in the form
\begin{equation}
\exp \bigl(-{\cal A}'(\chi)\bigr)= \int D\phi \prod_{blocks} f(\chi_{block};\phi)
\exp \bigl(-{\cal A}(\phi)\bigr) \, .
\label{genblock}
\end{equation}
Here $\phi, \chi$ can be scalar, vector or fermion fields and $f$ defines
how the coarse field in the block $\chi_{block}$ is constructed from the
(neighbouring) $\phi$ fields.

Assume, there is a symmetry transformation $\phi \rightarrow R\phi$ under
which the action and the measure are invariant: ${\cal A}(R\phi)={\cal A}(\phi)$,
$D(R\phi)=D\phi$. The blocked action ${\cal A}'(\chi)$ will inherit this symmetry,
${\cal A}'(R\chi)={\cal A}'(\chi)$, if the averaging function satisfies
\begin{equation} 
f(R\chi_{block};\phi)=f(\chi_{block};R^{-1}\phi)\, ,
\label{fc}
\end{equation}
as it is seen easily by changing integration variables in eq.~(\ref{genblock}).

In case of gauge symmetries we shall require somewhat more: we shall not only
require that the effect of a gauge transformation on the coarse lattice
is equivalent to a gauge transformation on the fine lattice, eq.~(\ref{fc}),
but also the other way around. This constraint implies that gauge equivalent
fine configurations contribute equally to the path integral in 
eq.~(\ref{genblock}).

\subsection{A basic assumption of the RG theory}

Consider a step of the RGT (eq.~(\ref{rgeq}) or eq.~(\ref{rgeqg})) for the case
where the parameters of ${\cal A}(\phi)$ were chosen so that the original system was
close to the continuum (or, in the language of statistical physics, close to
criticality). As eq.~(\ref{rgeqg}) shows, performing a RGT on a system with
action ${\cal A}(\phi)$ leads to a path integral where ${\cal A}$ is replaced by:
\begin{equation}
{\cal A}(\phi)+\kappa \sum_{n_B}\,\bigl( \chi_{n_B}-b\,\sum_{n}
\omega(2n_B-n) \phi_n \bigr)^2 \, .
\label{boltz}
\end{equation}
For this path integral the field $\chi$ enters as an external field. Even if 
${\cal A}(\phi)$ was carefully tuned to criticality, the system in eq.~(\ref{boltz})
is not expected to be critical. Actually, this is a basic assumption of the RG
theory: the path integral entering a step of the RGT defines a non-critical 
problem
with short-range interactions only. This is what one expects intuitively: the
$\chi$ fields constrain the $\phi$-averages in their neighbourhood and disrupt
the long-range fluctuations. This assumption leads to the important conclusion
that the new action ${\cal A}'(\chi)$ will be local. Really, if the r.h.s. of
eq.~(\ref{rgeqg}) describes a system with short-range fluctuations only, the
generated interaction between distant $\chi$ fields is expected to be
negligibly small.

Another consequence of this basic assumption is that the path integral in
eq.~(\ref{rgeq}) or eq.~(\ref{rgeqg}) is a technically much simpler problem
than the path integral of the original system. Saying differently: to perform
a RG step is not an easy problem in general, but it is much simpler than to
solve the original (near) critical theory. This is what makes the RG theory a
practically useful idea.

\subsection{Fixed Point (FP) and the behaviour \\
in the vicinity of a FP} 

In general, the transformed action ${\cal A}'(\chi)$ in eq.~(\ref{rgeqg}) will
contain all kinds of interactions, even if the original action ${\cal A}(\phi)$ had a
simple form. It is useful to introduce a sufficiently general interaction
space to describe the actions generated by the RGT, and write 
\begin{equation}
{\cal A}(\phi)= \sum_{\alpha} K_\alpha \theta_\alpha(\phi),
\label{theta}
\end{equation}
where $\theta_\alpha(\phi)$, $\alpha = 1,2,...$ denotes the different
interaction terms like (for scalar fields)
\begin{equation}
\theta_1=\sum_{n,\mu}\nabla_\mu \phi_n \nabla_\mu \phi_n \, ,\qquad
\theta_2=\sum_{n}\phi^2_n \; ,\ldots , \;  \theta_i=\sum_{n,\mu}
\phi^4_n \nabla_\mu \phi_n \nabla_\mu \phi_n ,\ldots
\label{terms}
\end{equation}
and $K_\alpha$ are the corresponding dimensionless couplings. The transformed
action ${\cal A}'(\chi)$ is also expanded in terms of these interaction terms
\begin{equation}
{\cal A}'(\chi)= \sum_{\alpha} K'_\alpha \theta_\alpha(\chi) \, .
\label{kprime}
\end{equation}
The RGT induces a motion in the coupling constant space:$\{ K_\alpha \}
\rightarrow \{ K'_\alpha \}$. Under repeated RGTs a coupling constant
flow is generated
\begin{equation}
\{ K_\alpha \} \rightarrow \{ K'_\alpha \} \rightarrow
\{ K''_\alpha \} \rightarrow \ldots ,
\label{flow}
\end{equation}
while the (dimensionless) correlation length is reduced with every step
(eq.~(\ref{xired})) 
\begin{equation}
\xi \rightarrow \frac{1}{2} \xi \rightarrow \frac{1}{4} \xi \rightarrow \ldots
\, .
\label{xiredrep}
\end{equation}

It might happen that certain points in the coupling constant space are
reproduced by the RGT
\begin{equation}
\{ K^*_\alpha \}  \rightarrow \{ K^*_\alpha \} \, .
\label{fp}
\end{equation}
A point with this property is called a fixed point (FP) of the RGT and the
corresponding action
\begin{equation}
{\cal A}^{FP}(\phi)= \sum_{\alpha} K^*_\alpha \theta_\alpha(\phi)
\label{sfp}
\end{equation}
is the FP action. FP actions play an important role in QFTs in general and
will play an important role in our later discussion of the perfect actions.

Eq.~(\ref{flow}) and eq.~(\ref{xiredrep}) imply that at the FP, 
$\xi =\infty$ or $\xi = 0$. We will be interested in FPs 
with $\xi=\infty$. The set of points in the coupling constant space where 
$\xi=\infty$, forms a hypersurface, which is called the critical surface. As   
eq.~(\ref{xiredrep}) shows, an RGT drives the point $\{ K_\alpha \}$ away
from the critical surface, except when $\{ K_\alpha \}$ is on the critical
surface.

Let us consider now the behaviour of the flow under the RGT in the 
vicinity of a FP \cite{weg}. Take a point $\{ K_\alpha \}$ with 
$\triangle K_\alpha = K_\alpha - K^*_\alpha$ small. Under the RGT we have
\begin{equation}
\triangle K_\alpha  \rightarrow \triangle K'_\alpha = K'_\alpha - K^*_\alpha
\, .
\label{deltak}
\end{equation}
Expanding $\triangle K'_\alpha = \triangle K'_\alpha(\{K\})$ around  $\{K^*\}$
\begin{equation}
\triangle K'_\alpha =\triangle K'_\alpha (\{ K^* \})+ \sum_{\beta}
\frac{\partial}{\partial K_\beta} \triangle K'_\alpha (\{ K \})_{\vert K=K^*}
\triangle K_\beta + O\bigl( (\triangle K)^2 \bigr)\, ,
\label{deltaexp}
\end{equation}
and observing that the first term on the r.h.s. is zero, we get the following 
linearized RG equation
\begin{equation}
\triangle K'_\alpha = \sum_{\beta} T_{\alpha \beta} \triangle K_\beta ,
\label{linrg}
\end{equation}
with 
\begin{equation}
T_{\alpha \beta} = 
\frac{\partial}{\partial K_\beta} \triangle K'_\alpha (\{ K \})_{\vert
  K=K^*} \, . 
\label{tmatrix}
\end{equation}
Let us denote the eigenvectors and eigenvalues of the matrix $T$ by $h_a$ and
$\lambda_a$, respectively:\footnote{Actually, the real matrix $T$ is not
symmetric in general. For a discussion on related consequences, see the
first reference in \cite{wil}}
\begin{equation}
\sum_\beta T_{\alpha \beta} h^a_\beta = \lambda^a h^a_\alpha \, ,\, a=1,2,
\ldots \, .
\label{eigen}
\end{equation}
The eigenvectors $h^a$ define the eigenoperators
\begin{equation}
h^a(\phi) = \sum_{\alpha} h^a_\alpha \theta_\alpha(\phi) \, .
\label{eigenop}
\end{equation}
These operators define a new basis in terms of which the action can be
expanded
\begin{equation}
{\cal A}(\phi) = {\cal A}^{FP}(\phi) + \sum_{a} c^a h^a(\phi) ,
\label{newexp}
\end{equation}
where $c^a$ are the corresponding expansion coefficients (couplings) which are
small if ${\cal A}$ is close to ${\cal A}^{FP}$. Repeated application of the RGT gives
\begin{equation}
{\cal A}(\phi) \rightarrow {\cal A}^{FP}(\phi)+\sum_{a}\lambda^a c^a h^a(\phi)
\rightarrow {\cal A}^{FP}(\phi)+\sum_{a}(\lambda^a)^2 c^a h^a(\phi) \rightarrow
\ldots \, ,
\label{sch}
\end{equation}
i.e., the coupling $c^a$ goes over to $(\lambda^a)^n c^a$ after $n$ RGT
steps. For $|\lambda^a| > 1$ ($|\lambda^a| < 1$) the coupling $|c^a|$ is 
increasing (decreasing) under repeated RG steps. The corresponding interaction
$h^a$ is called relevant (irrelevant). 
For $|\lambda^a| = 1$ the fate of the
operator (called marginal) is decided by the higher order corrections
suppressed in eq.~(\ref{deltaexp}).

\subsection{RG flows of a Yang-Mills theory}

The following discussion applies also for other AF theories, like the $O(N)$
non-linear $\sigma$-model or the $CP^n$ model in $d=2$.

The classical action of an $SU(N)$ Yang-Mills theory has the form
\begin{equation}
\beta {\cal A}_{cont} = \frac{\beta}{8N}\int_{x} F^a_{\mu \nu}(x)F^a_{\mu \nu}(x)
\, , \; a=1,\ldots,N^2-1\, ,
\label{contact}
\end{equation}
where $\beta=2N/g^2$ and $F^a_{\mu \nu}$ is the colour field strength
tensor. Any lattice representation of this action should go over to the form
in eq.~(\ref{contact}) for slowly changing weak fields. Even if we use a
gauge symmetric discretization (as we always do), there is a large freedom in
writing down a local action with this property on the lattice. As we discussed
in Sect.~3.4, we have to introduce an infinite dimensional coupling constant
space if we want to follow the change of the action under RGTs. We write
\begin{equation}
\beta {\cal A}_{latt} = \beta {\cal A}_{latt}(K_1,K_2,\ldots),
\label{alatt}
\end{equation}
where we indicated the coupling constant dependence of the action only. Under
a RGT we have: $\{ \beta,K_1,K_2,...\} \rightarrow \{ \beta',K'_1,K'_2,...\}$.

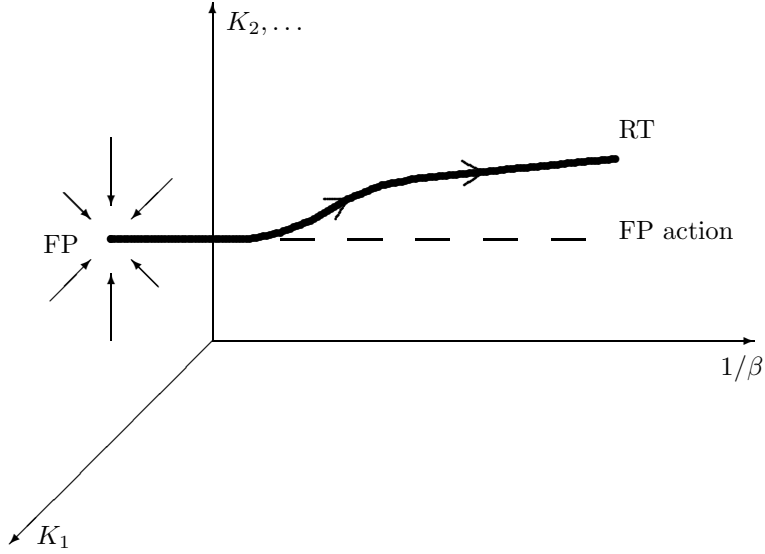
\begin{figure}  
\vskip1cm
\setlength{\unitlength}{.9mm}
\begin{center}
\begin{picture}(120,80)
\put(30,30){\vector(1,0){80}}
\put(30,30){\vector(0,1){50}}
\put(30,30){\vector(-1,-1){30}}
\put(4,0){$K_1$}
\put(32,76){$K_2,\ldots$}
\put(105,25){$1/\beta$}
\put(5,43){FP}
\put(15,45){\circle*{1}}
\put(15,30){\vector(0,1){10}}
\put(15,60){\vector(0,-1){10}}
\put(6,36){\vector(1,1){6}}
\put(24,54){\vector(-1,-1){6}}
\put(22,38){\vector(-1,1){4}}
\put(8,52){\vector(1,-1){4}}
\multiput(15,45)(.5,0){40}{\circle*{1}}
\multiput(35,45)(.5,.1){10}{\circle*{1}}
\multiput(40,46)(.5,.2){10}{\circle*{1}}
\multiput(45,48)(.5,.3){10}{\circle*{1}}
\multiput(50,51)(.5,.2){10}{\circle*{1}}
\multiput(55,53)(.5,.1){10}{\circle*{1}}
\multiput(60,54)(.5,0.05){60}{\circle*{1}}
\thicklines
\multiput(40,45)(10,0){5}{\line(1,0){5}}
\thinlines
\put(90,45){FP action}
\put(90,60){RT}
\multiput(50,51)(-.25,0){15}{\circle*{.5}}
\multiput(50,51)(-.13,-.21){15}{\circle*{.5}}
\multiput(70,55)(-.22,0.11){15}{\circle*{.5}}
\multiput(70,55)(-.20,-.14){15}{\circle*{.5}}
\end{picture}
\end{center}
\caption{Flow of the couplings under RG transformations
 in a Yang-Mills gauge theory.}
\end{figure}

The expected flow diagram is sketched in fig.~2. In the $\beta=\infty$
hyperplane, which is the critical surface, there is a FP with coordinates
$K^*_1,K^*_2,...$. We shall suppress the index `latt' and introduce the
notation ${\cal A}(K^*_1,K^*_2,...) = {\cal A}^{FP}$. All the eigenoperators which lie in
the $\beta =\infty$ hypersurface are irrelevant. There is one marginal
direction which is pointing out of this surface. Actually, this direction is
marginal in the linear approximation only, it becomes weakly relevant by the
higher order corrections to eq.~(\ref{deltaexp}).

Consider the point $\{ \beta,K_1,K_2,...\}$, $\beta$ very large. Under RGTs
this point runs rapidly towards the FP (the largest irrelevant eigenvalue is
1/4), then slowly moves away from it along the weakly relevant direction
tracing a trajectory, called the renormalized trajectory.

Consider the interaction corresponding to the FP, ${\cal A}^{FP}$, multiply it by
$\beta$: $\beta {\cal A}^{FP}$ and allow $\beta$ to move away from $\infty$. This is
a straight line (the dashed line in fig.~2.) which is not a RG flow, but
defines an action for every value of $\beta$. We shall call this action the FP
action.

We postpone the discussion of
the arguments supporting the structure of the
flow diagram in fig.~2. until Sect.~8.
We mention here only that theories with this flow
diagram can be renormalized by tuning a single coupling constant and the
resulting renormalized theory is universal (independent of the value of the
other couplings in the (bare) action). The continuum renormalized
theory is obtained as $g \rightarrow 0$ ($\beta \rightarrow \infty$) in which
limit the correlation length goes to infinity, the lattice unit $a \rightarrow
0$, the resolution becomes infinitely good and the cut-off artifacts disappear
from the predictions.

\section{Perfect actions in AF theories}

We shall call a lattice regularized local action \textit{classically perfect} if
its classical predictions (independently whether the lattice is fine or
coarse, whether the resolution is good or bad) agree with those of the
continuum. The \textit{quantum perfect action} does the same for all the physical
questions in the quantum theory. That such actions exist might seem to be
surprising. As we shall see, they do exist and have beautiful properties which
one would not expect a lattice action 
can have \cite{hn,has,sum,uwe1,bb1,mit1}.

We shall argue now that the actions defined by the points of the renormalized
trajectory (RT) in fig.~2 define quantum perfect actions. It means that by
taking an arbitrary point $(\beta,K_1,K_2,...) \in RT$, where $g^2=2N/\beta$
need not be small, the corresponding action ${\cal A}(\beta,K_1,K_2,...)$ will
produce quantum results for physical questions (mass ratios, for example)
which are exactly the same as in the continuum limit. This is surprising, since
in the point $(\beta,K_1,K_2,...)$ the correlation length is not large, the
lattice unit $a$ is not small, the resolution is not fine. Nevertheless, the
statement is true.

The argument goes as follows\cite{hn}. 
At any given $\beta$, the point of the RT is
connected to the infinitesimal neighbourhood of the FP by (infinitely many
steps of) exact RG transformations. Since each step increases the lattice unit
by a factor of 2, \textit{any} distance at the given $\beta$ (even 1 lattice unit)
corresponds to a long distance close to the FP. The infinitesimal
neighbourhood of the FP is in the continuum limit, there are no cut-off
effects there at long distances. On the other hand, for all the questions
which can be formulated in terms of the degrees of freedom after the
transformation we get the same answer as before the transformation. Thus,
there are no lattice artifacts at the given $\beta$ on the RT at any
distances.

We shall also argue that the FP action, as defined in Sect.~3.5 (fig.~2) is a
classically perfect action. We are not able to demonstrate this statement so
compactly as that for the quantum perfect action before. In the following
sections we shall go over the different classical properties of continuum
Yang-Mills, QCD and other AF theories and show one by one that they are
reproduced by the FP action independently of the lattice resolution.
An additional, unexpected property of the QCD FP action is, as we shall
discuss in some detail, that it reproduces
the physical consequences of chiral symmetry in the quantum theory exactly. 

The detailed form of the FP action and the RT depend on the form and
parameters of the block transformation. The theoretical properties of the FP
action and RT are, however, independent of these details. In this sense, any
RGT which has a local FP is suitable. On the other hand, different local FP actions
(belonging to different block transformations) have different extensions and
this has an important practical significance when it comes to parameterization
and simulation. In order to find short ranged actions (Sect.~2.4), different
block transformations were tested and 
optimized \cite{has}.

\section{Lattice regularization and explicit RGTs in different AF theories}  

It is time now to come down from generalities to concrete examples. We shall
discuss how the lattice regularization is set up and give explicit RGTs in
different AF theories. First, the $O(N)$ non-linear $\sigma$-model and the
$SU(N)$ Yang-Mills theory will be considered. Then, as a preparation for QCD,
the special difficulties of lattice regularized Dirac fermions will be
illustrated by treating the case of free spin-1/2 fermions. We close this
section with QCD.

In all these theories we write down the 'standard' lattice action which is the
simplest discretization of the continuum action satisfying the basic
requirements. These are the actions which - due to their technical simplicity
- were used in most of the numerical simulations. These actions are simple,
but they represent poorly most of the properties even of the classical field
theory if the resolution is not sufficiently fine. As we discussed in Sect.~4,
in the framework of the RG theory actions can be defined, which have
beautiful properties. For later use we define explicit RGTs in
all the theories considered in this section.

\subsection{O(N) non-linear $\sigma$-model in d=2}

The classical action in the continuum has the form
\begin{equation}
\beta {\cal A}_{cont}(\Sv) = {\beta\over 2} \int_{x}
\partial_{\mu} \Sv(x) \partial_{\mu} \Sv(x) \, ,
\label{sigcont}
\end{equation}
where $\Sv$ is an $N$-component vector satisfying the constraint $\Sv^2(x)=1$,
$\beta=1/g$. Writing
\begin{equation}
\Sv(x)=\left( {\sqrt{1-\pi(x)^2}
\atop  \pii(x)} \right)
\label{spi}
\end{equation}
and expanding in the fluctuations $\pii(x)$ 
(it has $N-1$ components), a
systematic perturbation theory can be set up. The theory is AF in the coupling $g$
for $N \geq 3$. It is believed (and this is strongly supported by different
theoretical and numerical results) that a non-zero mass is generated
dynamically, the spectrum contains a massive $O(N)$ multiplet. The
$\sigma$-model is discussed in different modern textbooks on QFT and
statistical systems \cite{text1,text2,text3}.

Looking for classical solutions with finite action, the spin vectors $\Sv$
should be parallel at large distances. For $N=3$ this leads to an $S_2
\rightarrow S_2$ mapping between the coordinate and group spaces. The $O(3)$
model has a non-trivial topology, there exist scale invariant instanton
solutions with an action ${\cal A} = 4\pi |Q|$, where $Q$ is the topological charge 
of the solution.

The $O(3)$ model shows many analogies \cite{mika,text3} with the 
Yang-Mills theory in $d=4$. It
offers the possibility to test non-perturbative ideas and numerical methods.

On the lattice, the scalar field $\Sv$ lives on the lattice points. The
simplest realization of the action has the form (`standard action')
\begin{equation}
\beta {\cal A}_{st}(\Sv) = {\beta\over 2}\sum_{n,\mu} 
\nabla_\mu \Sv_n \nabla_\mu \Sv_n \, ,
\label{sigst}
\end{equation}
where $\nabla_{\mu}$ is defined in eq.~(\ref{nnder}) and $\Sv_n^2=1$. On
smooth configurations ${\cal A}_{st}$ goes over to ${\cal A}_{cont}$ -- a property we shall
require for any lattice action.

Consider RGTs with a scale factor of 2, forming averages out of the four fine
spins in a block. The block spin will be denoted by $\R_{n_B}$,
$\R_{n_B}^2=1$. The RGT has the form
\begin{equation}
\exp \bigl(-\beta'{\cal A}'(\R)\bigr) = \int D\Sv 
\exp \bigl[-\beta \bigl( {\cal A}(\Sv) + T_\sigma(\R,\Sv) \bigr)\bigr] \, ,
\label{sigrgt}
\end{equation}
where $T_{\sigma}$ defines the averaging process and we used the notation
\begin{equation}
\int D\Sv = \prod_{n} \int d\Sv_n \delta(\Sv^2 - 1) \, .
\label{sigmeas}
\end{equation}

We shall give two explicit examples for the averaging which
were studied in the context of perfect actions \cite{hn,hn1}. 
Both RGTs are $O(N)$ invariant,
satisfying eq.~(\ref{fc}) for global $O(N)$ transformations.
In the first case:
\begin{equation}
T_{\sigma 1}(\R,\Sv)= \sum_{n_B}\bigl( -\kappa_\sigma \R_{n_B}\Si_{n_B} + 
{\cal N}(| \Si_{n_B} |) \bigr) \, ,
\label{sigt1}
\end{equation}
where $\Si_{n_B}$ is the sum of the four $\Sv$ spins in the block $n_B$
\begin{equation}
\Si_{n_B} = \sum_{n \in n_B} \Sv_n \, ,
\label{sigavr1}
\end{equation}
and ${\cal N}$ assures the correct normalization, eq.~(\ref{zpez}). We get
\begin{equation}
{\cal N}(| \Si_{n_B} |) = {1 \over \beta} 
\ln Y_N(\beta \kappa_\sigma | \Si_{n_B} |) \, ,
\label{signorm}
\end{equation}
where $Y_N(z)$ is related to the modified Bessel function (some of its
properties are summarized in appendix F in \cite{leut}, 
for example), specifically
$Y_3(z) \sim \sin(z)/z$. For $\kappa_\sigma \rightarrow \infty$, the block transformation
goes over to a $\delta$-function constraining the average to lie parallel to
the coarse spin
\begin{equation}
\frac{\Si_{n_B}}{|\Si_{n_B}|} = \R_{n_B}, \qquad (\kappa_\sigma=\infty) \, .
\end{equation}
It will be useful, however, to keep $\kappa_\sigma$ finite and optimize its value to
get a short ranged FP action.

Modifying slightly the averaging function $T$
\begin{equation}
T_{\sigma 2} = 2\kappa_\sigma \sum_{n_B} \bigl( \R_{n_B} - 
\frac{\Si_{n_B}}{|\Si_{n_B}|}\bigr)^2 \, ,
\label{sigt2}
\end{equation}
we can make the technically cumbersome normalizing function ${\cal N}$
constant. Really, the integral
\begin{equation}
\int d\R \, \delta(\R^2-1) \exp(4\beta \kappa_\sigma {\bf e} \R )\, ,
\end{equation}
with ${\bf e} = \Si/| \Si|$, depends on the length of
${\bf e}$ only, therefore it is an $\Sv$-independent constant.

\subsection{SU(N) Yang-Mills theory in d=4}

The classical action in the continuum has the form
\begin{equation}
\beta {\cal A}^{YM}_{cont}(A_{\mu}) = {\beta \over 4N} \int_{x} 
Tr(F_{\mu \nu}F_{\mu \nu}) \, ,
\label{ymcont}
\end{equation}
where $\beta=2N/g^2$, $F_{\mu \nu} = F^a_{\mu \nu}T^a, a=1,...,N^2-1$ and
$T^a$ are the generators of the colour group $SU(N)$, $Tr(T^a T^b) = 1/2
\delta_{ab}$.

\noindent The classical theory has scale invariant instanton 
solutions\cite{text4} 
with an action
${\cal A} = 8 \pi^2|Q|$.

The quantum theory is AF in the coupling $g$. It is believed (and this is
supported by many, but non-rigorous results) that non-perturbative effects
create a confining potential between static sources in the fundamental colour
representation (static quarks) and the spectrum contains massive colour
singlet excitations (glueballs) only.

On the lattice, the vector gauge field lives on the links of the hypercubic
lattice \cite{gaul,mont}. 
The convenient variables, in terms of which gauge invariance is
easily kept on the lattice are not the vector potentials themselves, but their
exponentiated forms. Consider a static quark and antiquark in the points $x$
and $y$, respectively (in the continuum). The form
\begin{equation}
\bar{\psi}(x)\, {\cal P} \exp\bigl(i\int_{x}^{y} A_\mu(z)dz_\mu\bigr)\, 
\psi(y) \, ,
\label{qfluxq}
\end{equation}
where the integral runs along some path between $x$ and $y$ and ${\cal P}$
denotes path ordering, is gauge invariant. The gauge part in
eq.~(\ref{qfluxq}), which leads the colour flux between the sources, is an
element of the $SU(N)$ group rather than that of the algebra. Going to the
lattice and placing $x,y$ in the endpoints of the link $(n,n+\hat \mu)$, we
get
\begin{equation}
\bar{\psi}_n\, U_\mu(n)\, \psi_{n+\hat{\mu}} \, ,
\label{qfluxqlatt}
\end{equation}
where $U_{\mu}(n)$ is related to the lattice vector potential as
\begin{equation}
U_{\mu}(n) = \exp(i a A_\mu(n)) \, .
\label{ua}
\end{equation}
The variables of the lattice regularized Yang-Mills theory are the link
matrices $U_{\mu}(n)$. The link matrix associated with the same link but
directed oppositely: $(n+\hat \mu,n)$ is $U^\dagger_{\mu}(n)$. Under a gauge
transformation the link matrix transforms as
\begin{equation}
U_{\mu}(n) \rightarrow V(n) U_{\mu}(n) V^\dagger(n+\hat{\mu}) \, , 
\label{gautr}
\end{equation} 
where $V(n) \in SU(N)$ is the gauge transformation in the point
$n$. Consequently,
the trace of the product of link matrices along a closed path on the lattice
is gauge invariant.

The simplest lattice realization of the Yang-Mills action has the form
(`Wilson action')
\begin{equation}
\beta {\cal A}_W(U) = {\beta \over N}\sum_{plaq}(N-ReTr\,U_p) \, ,
\label{wilact}
\end{equation}
where the sum is over the plaquettes and $U_p$ is the product of the four directed
link matrices around the plaquette $p$. For smooth gauge fields
eq.~(\ref{wilact}) goes over to eq.~(\ref{ymcont}), as it should.

We shall consider RGTs with a scale factor of 2. The blocked link variable
$V_{\mu}(n_B)\in SU(N)$, which lives on the coarse lattice with lattice unit
$a' = 2a$, is coupled to a local average of the original link variables. The
blocking kernel $T(V,U)$ enters the RGT as in the $\sigma$-model
\begin{equation}
\exp \bigl(-\beta'{\cal A}'(V)\bigr) = \int DU 
\exp \bigl[-\beta \bigl( {\cal A}(U) + T_g(V,U) \bigr) \bigr]\, ,
\label{ymrgt}
\end{equation}
where
\begin{equation}
\int DU = \prod_{n,\mu} \int dU_\mu(n) \, .
\label{ymmeas}
\end{equation}
The kernel $T_g$ is taken in the form
\begin{equation}
T_g(V,U) = \sum_{n_B,\mu} \bigl[-{\kappa_g \over N}
ReTr\,\bigl( V_\mu(n_B) Q^\dagger_\mu(n_B) \bigr) + 
{\cal N}\bigl( Q_\mu(n_B) \bigr) \bigr] \, .
\label{tym}
\end{equation}
The $N \times N$ complex matrix $Q_{\mu}(n_B)$ is an average of the fine link
variables in the neighbourhood of the coarse link $l_B=(n_B,\mu)$. 
${\cal N}(Q_{\mu}(n_B))$ is fixed by the normalization condition, eq.~(\ref{zpez}):
\begin{equation}
\exp\bigl({\cal N}(Q)\bigr) = \int dW \exp\bigl(\frac{\beta \kappa_g}{N}
ReTr\,(WQ^\dagger)\bigr), \qquad W \in SU(N) \, .
\label{ymn}
\end{equation}
Let us give a simple example for the construction of the average link matrix $Q$
\cite{swen,bb1} (fig.~3):
\begin{multline}
Q_\mu(n_B)  =  (1-6c)U_\mu(n)U_\mu(n+{\hat \mu}) + \\
c \sum_{\nu \ne \mu} \left[
U_\nu(n)U_\mu(n+{\hat \nu})U_\mu(n+{\hat \mu}+{\hat
  \nu})U^\dagger_\nu(n+2{\hat \mu})  + \right .\\
\left. U_\nu^\dagger(n-{\hat \nu})U_\mu(n-{\hat \nu})U_\mu(n+{\hat \mu}-{\hat
  \nu})U_\nu(n+2{\hat \mu}-{\hat \nu})
 \right] \, ,
\label{rgta}
\end{multline}
where $c$, the relative weight of the staples versus the central link, is a
tunable parameter (see, last paragraph of Sect.~4).

\begin{figure}[htb]
\begin{center}
\vskip 2mm\hskip-1cm
\leavevmode
\epsfxsize=80mm
\epsfbox{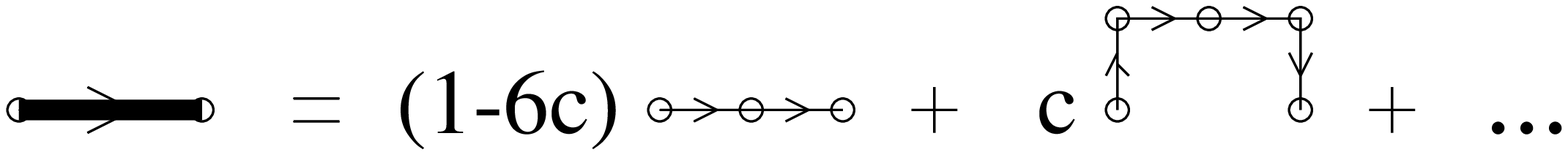}
\vskip-2mm
\end{center}
\caption{A simple example for a gauge invariant blocking}
\label{fig:type1}
\end{figure}

The averaging process in eq.~(\ref{rgta}) is not a very good one: there are
many link variables on the fine lattice which do not contribute to any of the
$Q$'s. Experience shows that poor averaging might lead to a FP which is
local, but not sufficiently short ranged or might not even have a FP. It is
not difficult, however, to generalize the block transformation above and to
make it physically more satisfactory \cite{mf}.

\subsection{Free fermion fields in d=4. Chiral symmetry, doubling and no-go
  theorems} 

The action of spin 1/2 Dirac fermions in the continuum has the form
\begin{equation}
{\cal A}_{cont}(\bar{\psi},\psi) = \int_{x} 
\bar{\psi}(x)(\gamma_\mu \partial_\mu +m) \psi(x) \, ,
\label{fcont}
\end{equation}
where $\{ \gamma_{\mu},\gamma_{\nu}\}=2\delta_{\mu \nu}$
and $\gamma_5=\gamma_1 \gamma_2 \gamma_3 \gamma_4$. 

On the lattice, the fermion field lives on the lattice points. The most
general action which is invariant under cubic rotation, reflection,
permutation of the coordinate axes and discrete translations, reads
\begin{equation}
{\cal A}(\bar{\psi},\psi) = \sum_{n,n'}\,\bar{\psi}_n h(n-n') \psi_{n'} \, , 
\label{fgen}
\end{equation}
where
\begin{equation}
h(n-n')= \rho_\mu(n-n')\gamma_\mu + \lambda(n-n') \, . 
\label{fgenx}
\end{equation}
The local functions $\rho_\mu$ and $\lambda$
have the following reflection properties
\begin{equation}
\rho_\mu(-r) =- \rho_\mu(r),\qquad \lambda(-r)=\lambda(r)\,,
\label{frefl}
\end{equation} 
and so $h$ satisfies
\begin{equation}
h^\dagger = \gamma_5 h \gamma_5 \,.
\label{herm}
\end{equation} 
A naive discretization of eq.~(\ref{fcont}) leads to
\begin{equation}
{\cal A}_{naive}(\bar{\psi},\psi) = {1 \over 2}\sum_{n,\mu}\,
(\bar{\psi}_n\gamma_\mu \psi_{n+\hat{\mu}}-
\bar{\psi}_{n+\hat{\mu}}\gamma_\mu \psi_n) +
\sum_{n}\,m \bar{\psi}_n \psi_n \, , 
\label{anaive}
\end{equation}
corresponding to
\begin{equation}
\rho_\mu(r)=\frac{1}{2}(\delta_{r+\hat{\mu},0}-\delta_{r-\hat{\mu},0}), \qquad 
\lambda(r)=m \delta_{r,0} \;,
\label{fnaive}
\end{equation}
or, in Fourier space
\begin{equation}
\rho_\mu(k)=i\sin(k_{\mu}), \qquad \lambda(k)=m \;.
\label{fnaivek}
\end{equation}
The momentum space propagator has the form
\begin{equation}
D(k) = \frac{-i \sum_{\mu}\, \sin(k_\mu)\gamma_\mu + m}
{\sum_{\mu}\, \sin^2(k_\mu) + m^2} \,.
\label{propnaive}
\end{equation}
The poles of the denominator determine the spectrum. Since $\sin(k_{\mu})=0$
at $k_{\mu}=0$ and $k_{\mu}=\pi$, the propagator in eq.~(\ref{propnaive}) has
16 poles defining 16 fermion species with the correct relativistic dispersion
relation $E=-ik_4=(\vec{k}^2 + m^2)^{1/2}$ in the continuum limit. This is the
species doubling problem specific to fermions on the lattice.

For $m=0$, the action in eq.~(\ref{fcont}) is chiral invariant, i.e. it
remains unchanged under the transformation
\begin{equation}
\psi(x) \rightarrow \exp(i \alpha \gamma_5)\, \psi(x) \, , \qquad
\bar{\psi}(x) \rightarrow \bar{\psi}(x)\, \exp(i \alpha \gamma_5)\,.
\label{cht}
\end{equation}
This symmetry requires on the lattice
\begin{equation}
\{h,\gamma_5 \}=0 \,,
\label{chirnaive}
\end{equation}
where we denoted the anticommutator by $\{,\}$. The naive action 
eq.~(\ref{anaive})
satisfies eq.~(\ref{chirnaive}) for $m=0$. In general, eq.~(\ref{chirnaive})
require $\lambda=0$ in eq.~(\ref{fgenx}) and leads to a propagator
$(\sum_\mu \rho_\mu\gamma_{\mu})^{-1}$. For small $k$, $\rho_\mu(k) \sim i
k_{\mu}$
since the lattice action should reproduce the continuum action in this
limit. In addition, $\rho_\mu(k)$ is a periodic function (Sect.~2.2). It
follows then that $-i \rho_\mu(k) > 0$ for $k_{\mu}$ small and positive and $<0$
for $k_{\mu}$ close to, but less than $2\pi$. If $\rho_\mu(k)$ is a continuous
function, it should have a zero in between. This intuitive argument implies a
close relation between the chiral symmetry of the action and the doubling
problem: if the action is chiral symmetric, we 
have species doubling \cite{nn}. The no-go theorem says: if the
lattice action has the correct continuum limit, i.e. $h(k)=i\gamma_\mu k_\mu$
for small momenta and there are no other zeros in the Brillouin zone (no
doublers), then one of the following two conditions should be violated:
a/ $h(k)$ is an analytic periodic function of $k_\mu$,
b/ $\{h,\gamma_5 \}=0$.
Clearly, we do not want to violate condition a/ because this would imply
loosing locality.
   
So, we have to violate the chiral symmetry of the action. It is well known
in perturbation theory in models without chiral symmetry (scalar QFT) or when
non-chiral invariant regularization is used (QED, for example, with
Pauli-Villars regularization) that radiative corrections induce an additive
mass renormalization and so, a tuning problem. This will be the generic
situation in our case also. Similarly, the axial current will not be
conserved, the soft pion theorems will be violated and there will occur
mixing between operators in different nominal chiral representations.
However, as we discuss later, the FP action offers an
elegant solution to all these problems.

A simple way to kill the doublers is to add to the action an appropriate
non-chiral invariant term which disappears in the continuum limit. Wilson
suggested \cite{wf} (`Wilson fermion action') 
to add to eq.~(\ref{anaive})
\begin{equation}
{r \over 2}\sum_{n,\mu}\,\bigl( 2 \bar{\psi}_n \psi_n -\bar{\psi}_n
\psi_{n+\hat{\mu}}-
\bar{\psi}_{n+\hat{\mu}} \psi_n \bigr)
\label{wterm}
\end{equation}
with $r\ne 0$ a free parameter. This term contributes to $\lambda(k)$
\begin{equation}
\lambda(k) = m + 2r\sum_{\mu} \sin^2({k_\mu \over 2})\, .
\label{fwilk}
\end{equation}
The extra term gives an $O(1/a)$ mass to the doublers, but it is suppressed
around the $k \sim 0$ pole. This procedure can be generalized to interactive
theories also and this is the method which is used in most of the QCD
calculations.

Consider now RGTs on a $d$-dimensional hypercubic lattice with a scale factor
of 2. Write the RGT in the form
\begin{equation}
\exp \bigl( -{\cal A}'(\bar{\chi},\chi) \bigr) = \int D\bar{\psi}D\psi
\exp \bigl[ -\bigl( {\cal A}(\bar{\psi},\psi)+ 
T_f(\bar{\chi},\chi;\bar{\psi},\psi) \bigr) \bigr]
\label{frgt}
\end{equation}
with
\begin{equation}
\int D\bar{\psi}D\psi = \prod_{n} \int d\bar{\psi}_nd\psi_n \, ,
\label{fmeas}
\end{equation}
where $\bar{\psi}, \psi$ are Grassmann variables. Consider blocking kernels of
the form
\begin{multline}
T_f(\bar{\chi},\chi;\bar{\psi},\psi) =
\kappa_f \sum_{n_B}\,\bigl( \bar{\chi}_{n_B}-b_f \sum_{n}\,\bar{\psi}_n
\omega(2n_B-n) \bigr) \\
\times\bigl( \chi_{n_B}-b_f \sum_{n}\,\omega(2n_B-n) \psi_n \bigr)\,.
\label{tfer}
\end{multline}
We shall assume that the averaging function $\omega$ is diagonal in Dirac
space and real. Observe that $T_f$ is not chiral invariant.

We shall consider two different block 
transformations \cite{uwe1,bb2,ku}. 
The first is just the
blocking introduced for a scalar field in Sect.~3.1 and also for the
$\sigma$-model in Sect.~5.1: $\omega(2n_B-n)=2^{-d}$ if $n$ is in
the hypercube whose center in indexed by $n_B$ and zero otherwise.

Keeping in mind QCD, the blocking in the second example has a form
which is easy to make gauge invariant in the presence of gauge interactions.
Place the block fields in the even points of the
fine lattice ($n$ is even if all its coordinates are even). A simple choice
for the averaging function $\omega$ in this case is
\begin{multline}
\omega(0,0,\ldots,0)=c_0,\qquad 
\omega(\pm 1,0,\ldots,0)=\ldots =\omega(0,0,\ldots,\pm 1)=c_1,\\
\ldots,
\omega(\pm 1,\pm 1,\ldots,\pm 1) = c_d \,.
\label{omc}
\end{multline}
The parameters $c_0,c_1,...,c_d$ satisfy the normalization condition, eq.~(\ref{omnorm}),
\begin{equation}
c_0 +2d\, c_1 + 2d(d-1)\,c_2+ \ldots + 2^d \,c_d = 1\, .
\label{normc}
\end{equation}
Observe that the fine field $\psi_n$ contributes to several block averages, in
general. We shall call a blocking with this property 'overlapping'. The block
transformations in eqs.~(\ref{sigt1},\ref{sigt2}) and the first example above 
are 'non-overlapping', while those
in eqs.~(\ref{rgta},\ref{omc}) are overlapping.

There is a natural symmetric choice for the $c_i$ parameters in
eq.~(\ref{omc}):
\begin{equation}
c_i= 2^{-d-i},\qquad i=0,1,...,d \,.
\label{cflat}
\end{equation}
In this case, the sum of contributions of $\psi_n$ to all the block fields
$\chi$ is independent of $n$, the averaging is `flat'. 

\subsection{QCD}

The action is the sum of the Yang-Mills action and the action of quarks in
interaction with the colour gauge field
\begin{equation}
{\cal A}^{QCD}_{cont}(\bar{\psi},\psi,A_{\mu}) = \beta {\cal A}^{YM}_{cont}(A_\mu)
+\int_{x} \bar{\psi}(x) \bigl[ \bigl( \gamma_\mu
(\partial_\mu-iA_\mu(x) \bigr)+m \bigr] \psi(x)\,.
\label{qcdcont}
\end{equation}
The quark fields carry Dirac, colour and flavour indices, the gauge field does
not know about flavour, the mass (matrix) does not know about colour. For
$N=3$, this theory is believed to be the theory of the hadrons. It should
explain the masses, widths and other static properties of hadrons, scattering
events, all kinds of hadronic matrix elements, the behaviour of hadronic matter
at finite temperature and density, spontaneous chiral symmetry breaking, the
$\eta'$ problem, etc. Most of these problems are non-perturbative.

The simplest lattice regularized QCD action, which we shall call the `Wilson
action' is the sum of the gauge action eq.~(\ref{wilact}) and the fermionic
action, eq.~(\ref{anaive}) plus eq.~(\ref{wterm}), made gauge invariant. Take
$r=1$, which is the preferred value (for technical reasons) in simulations
\begin{multline}
\A^{QCD}_W(\bar{\psi},\psi,U) = {\beta \over N} \sum_{plaq}(N-ReTr\,U_p) + \\
{1 \over 2}\sum_{n,\mu}
\bigl( \bar{\psi}_n(\gamma_\mu-1)U_\mu(n) \psi_{n+\hat{\mu}}-
\bar{\psi}_{n+\hat{\mu}}(\gamma_\mu+1)U^\dagger_\mu(n) \psi_n \bigr) 
+\sum_{n}\bar{\psi}_n(m+4)\psi_n \, . 
\label{qcdw}
\end{multline}

\noindent We write the RGT in the form
\begin{multline}
\exp\bigl[-\bigl( \beta' \A'_g(V)+\A'_f(\bar{\chi},\chi,V)\bigr) \bigr]= \\
\int D\bar{\psi}D\psi DU \exp \bigl[ -\beta \bigl(\A_g(U)+T_g(V,U)\bigr)+
\bigl(\A_f(\bar{\psi},\psi,U)+T_f(\bar{\chi},\chi;\bar{\psi},\psi,U) \bigr)
\bigr]\, . 
\label{qcdrgt}
\end{multline}
The gauge kernel $T_g$ is defined in eq.~(\ref{tym}), while the fermion kernel
is the gauge invariant version of eq.~(\ref{tfer})
\begin{multline}
T_f(\bar{\chi},\chi;\bar{\psi},\psi,U) =
\kappa_f \sum_{n_B}\,\bigl( \bar{\chi}_{n_B}-b_f \sum_{n}\,\bar{\psi}_n
\omega^\dagger(U)_{n,n_B} \bigr) \\
\times\bigl( \chi_{n_B}-b_f \sum_{n}\,\omega(U)_{n_B,n} \psi_n \bigr)\,.
\label{tfqcd}
\end{multline}
In eq.~(\ref{tfqcd}), the averaging function $\omega$ is made gauge invariant
by connecting $n_B$ with the point $n$ by a product of $U$ matrices. This can
be done without problems if $n_B$ overlaps with one of the fine lattice points
as in the second example in Sect.~5.3, eq.~(\ref{omc}). Beyond
hypercubic symmetry the choice of these paths is part of the freedom we have
in defining an RGT. A simple choice is to take the shortest connecting
paths. For $U \rightarrow 1$, $\omega(U)_{n_B,n} \rightarrow \omega(2n_B-n)$
defined in eq.~(\ref{tfer}), which satisfies the normalization condition,
eq.~(\ref{omnorm}). This condition gives a meaning to $b_f$ in
eq.~(\ref{tfqcd}). As we shall see (Sect.~8), in a free field theory $b_f$ is
determined by the engineering dimension of the field. This should be the case
also in the classical theory even in the presence of interaction, since the
corrections to $b_f$ are related to the anomalous dimension of the field
generated by quantum fluctuations \cite{wil,others,text1}. 
We shall return to this point later.

\section{The saddle point equation for the fixed point action} 

For free field theories with a quadratic blocking kernel, like those in
eq.~(\ref{tfer}) or in eq.~(\ref{rgeqg}), the RGT  is
reduced to Gaussian integrals. The integrals can be performed exactly and the
problem can be investigated in every detail. In general, however, the
path integral of a RGT is a highly non-trivial problem.

A basic observation to proceed is that in AF theories the FP lies at
$\beta=\infty$ and in this limit the path integral can be calculated in the
saddle point approximation \cite{hn}. 
This leads to an equation in classical field
theory which determines the FP action. We shall derive this
equation first in the non-linear $\sigma$-model and then in QCD.

\subsection{The FP of the O(N) non-linear $\sigma$-model}

For a
given fixed coarse field configuration $\R$ we can perform the path integral
\begin{equation}
\int D\Sv \exp \bigl[-\beta \bigl( {\cal A}(\Sv) + T_\sigma(\R,\Sv) \bigr) \bigr]
\label{tdefs}
\end{equation}
in the $\beta \to \infty$ limit
using the saddle point approximation leading to the classical equation
\begin{equation}
{\cal A}'(\R) =\min_{\{\Sv \}} \bigl[ {\cal A}(\Sv) + T^\infty_\sigma(\R,\Sv) \bigr]\, ,
\label{mins}
\end{equation}
where $T^\infty_\sigma$ is the $\beta=\infty$ limit of the blocking kernel.
The FP of the transformation is determined by the equation
\begin{equation}
{\cal A}^{FP}(\R) =\min_{\{\Sv \}} \bigl[ {\cal A}^{FP}(\Sv) + 
T^\infty_\sigma(\R,\Sv) \bigr]\, .
\label{fps}
\end{equation}
Let us write down this equation explicitly for the block transformation defined in
eqs.~(\ref{sigt1},\ref{signorm}):
\begin{equation}
{\cal A}^{FP}(\R) =\min_{\{\Sv \}} \bigl[ {\cal A}^{FP}(\Sv) 
-\kappa_\sigma \sum_{n_B}\bigl( \R_{n_B}\Si_{n_B} - |\Si_{n_B}| \bigr)
\bigr]\, ,
\label{fps1}
\end{equation}
where we have used $\ln Y_N(z)=z (1+O(\ln z/z))$ for large $z$ and $\Si_{n_B}$
is defined in eq.~(\ref{sigavr1}).

Observe that eq.~(\ref{tdefs}) is reduced to the saddle point equation,
eq.~(\ref{mins}), for any configuration $\R$. If the configuration $\R$ is
strongly fluctuating then the minimizing configuration $\Sv_{min}$ will not be
smooth either. In general, eqs.~(\ref{mins},\ref{fps}) and their
solutions have nothing to do with perturbation theory. A starting condition
for the FP equation, eq.~(\ref{fps}), is that ${\cal A}^{FP}$
goes over to the classical action ${\cal A}_{cont}(\R)$, eq.~(\ref{sigcont}), for very
smooth $\R$ configurations.  

\subsection{The FP of QCD}

In the $\beta \rightarrow \infty$ limit ($V, \bar{\chi}, \chi$ fixed), the
Boltzmann factor on the r.h.s. of eq.~(\ref{qcdrgt}) is dominated by the gauge
part. The integral over the $U$ field is saturated by the minimizing
configuration $U_{min}=U_{min}(V)$ leading to the equation
\begin{equation}
{\cal A}_g'(V) =\min_{\{U \}} \bigl[ {\cal A}_g(U) + T^\infty_g(V,U) \bigr]\, .
\label{minym}
\end{equation}
The gauge part of the FP is determined by
\begin{equation}
{\cal A}^{FP}_g(V) =\min_{\{U \}} \bigl[ {\cal A}^{FP}_g(U) + T^\infty_g(V,U) \bigr]\, .
\label{fpym}
\end{equation}
Using the explicit form of the blocking kernel $T_g$ in eq.~(\ref{tym}) we
obtain
\begin{equation}
{\cal A}^{FP}_g(V) = 
\min_{\{U \}} \left( {\cal A}^{FP}_g(U)
-{\kappa_g \over N} \sum_{n_B,\mu}\,\bigl[ ReTr\bigl(V_\mu(n_B)Q^\dagger_\mu(n_B)\bigr)-
f\bigl(Q_\mu(n_B)\bigr) \bigr] \right),
\label{fpyme}
\end{equation}
where
\begin{equation}
f(Q) = \max_{W} \bigl[ ReTr(WQ^\dagger) \bigr],\qquad W \in SU(N) \, .
\label{fpymnorm}
\end{equation}
Eq.~(\ref{fpymnorm}) follows from eq.~(\ref{ymn}) in the $\beta \rightarrow
\infty$ limit.

The remaining fermionic integral has the form
\begin{equation}
\exp \bigl( -{\cal A}'_f(\bar{\chi},\chi,V) \bigr) = \int D\bar{\psi}D\psi
\exp \bigl[ -\bigl( {\cal A}_f(\bar{\psi},\psi,U_{min})+ 
T_f(\bar{\chi},\chi;\bar{\psi},\psi,U_{min}) \bigr) \bigr],
\label{gaussf}
\end{equation}
where $U_{min}(V)$ is the minimizing configuration from eq.~(\ref{minym}). If the
action ${\cal A}_f$ is quadratic in the fermion fields, then the blocked action
${\cal A}'_f$ is quadratic also, since the kernel $T_f$ is quadratic
(eq.~(\ref{tfqcd})) and the integral in eq.~(\ref{gaussf}) is Gaussian. Write
the fermion action in the following general form
\begin{equation}
{\cal A}_f(\bar{\psi},\psi,U) = \sum_{n,n'}\bar{\psi}_n h_{n,n'}(U)\psi_n'\, ,
\label{hdef}
\end{equation}
where $h(U)_{n,n'}$ carries Dirac, colour and flavour indices which are 
not indicated
explicitly. Using eq.~(\ref{tfqcd}) and performing the Gaussian integral in
eq.~(\ref{gaussf}) we get the following recursion relation for $h$
\begin{eqnarray}
\label{27}
& & h(V)_{n_B n_B'} = \kappa_{\rm f}\delta_{n_B n_B'} 
~~~~~~~~~~~~~~~~~~~~~~~~~~~~~~~~~~~~~~~~~~~~~~~~~~~~~~~~~~~~~~~~ \\
& & ~~~ - \kappa_{\rm f}^2 b_{\rm f}^2 \sum_{nn'}
\omega(U_{\rm min})_{n_B n} \left( h(U_{\rm min}) +   
\kappa_{\rm f} b_{\rm f}^2 \omega^{\dagger}(U_{\rm min})
\omega(U_{\rm min}) \right)^{-1}_{n n'}
\omega(U_{\rm min})_{n' n_B'}^{\dagger} \,. \nonumber
\end{eqnarray}
In case $h^{-1}$ is defined (no zero modes, see Sect.~10) eq.~(\ref{27}) is
equivalent to the somewhat simpler equation \cite{bb2,wuj}
\begin{equation}
h'(V)^{-1}_{n_B,n'_B} = 
{1 \over \kappa_f}\delta_{n_B,n'_B} +
b^2_f \sum_{n,n'}\omega(U_{\rm min})_{n_B,n}h(U_{\rm min})^{-1}_{n,n'}
\omega(U_{\rm min})^\dagger_{n',n'_B} \, .
\label{transfh}
\end{equation}
The FP satisfies the equation
\begin{equation}
h^{FP}(V)^{-1}_{n_B,n'_B} = {1 \over \kappa_f}\delta_{n_B,n'_B} +
b^2_f \sum_{n,n'}\omega(U_{\rm min})_{n_B,n}h^{FP}(U_{\rm min})^{-1}_{n,n'}
\omega(U_{\rm min})^\dagger_{n',n'_B} \, .
\label{fph}
\end{equation}
Eqs.~(\ref{fpyme},\ref{fph}) determine the FP action of QCD. Gaussian integrals
of c-number fields are equivalent to minimization. This can be generalized to
Gaussian Grassmann integrals to express eq.~(\ref{fph}) as
\begin{equation}
{\cal A}^{FP}_f(\bar{\chi},\chi,V) = \bigl[ {\cal A}^{FP}_f(\bar{\psi},\psi,U_{\rm min})+
T_f(\bar{\chi},\chi;\bar{\psi},\psi,U_{\rm min})\bigr]_{|\bar{\psi}_{st},
\psi_{st}}
\label{fph1}
\end{equation}
where $\psi_{st},\bar{\psi}_{st}$ make the r.h.s. stationary, 
$\delta/\delta\psi$~[...]=0, $\delta/\delta\bar{\psi}$~[...]=0. In
eq.~(\ref{fph1}) the fields $\bar{\chi},\chi,\bar{\psi},\psi$
are c-number fields.

\section{The FP action for weak fields}

We have claimed in Sect.~4 that the FP action is classically perfect: it
reproduces all the essential physical properties of the continuum classical
action on the lattice even if the resolution of the lattice is poor. The proof
of this statement follows directly from the FP equations (eq.~(\ref{fps1}) for
the $\sigma$-model or eqs.~(\ref{fpyme},\ref{fph}) for QCD) without solving
them explicitly. Before presenting these arguments let us study the FP
equations in a limit where analytic results can be obtained. 

Consider first the FP equation in the $O(N)$ non-linear $\sigma$-model,
eq.~(\ref{fps1}). Take a configuration where the spins fluctuate around the
first axis in $O(N)$ space
\begin{equation}
\R_{n_B}=\left( {\sqrt{1-{\boldsymbol \chi}^2_{n_B}}
\atop {\boldsymbol \chi}_{n_B}} \right)\, ,
\label{r1ax}
\end{equation}
where ${\boldsymbol \chi}_{n_B}$ 
has $(N-1)$ components. Assume, these fluctuations are
'weak', $|{\boldsymbol \chi}_{n_B}| \ll 1$. 
In this case, the minimizing field $\Sv_{min} (\R)$
will also fluctuate around the first axis, eq.~(\ref{spi}), and these
fluctuations will also be weak, $|{\boldsymbol \pi}_{n_B}| \ll 1$. For weak fields
eq.~(\ref{fps1}) can be expanded in powers of the fluctuations and one can study
the solution order by order.

There is a difference between the 'weak field' introduced here and the 'smooth
field' used repeatedly earlier. A smooth field changes slowly in
configuration space and is dominated by small $k$ values in Fourier space. In a
weak field, the size of deviations from the classical vacuum configuration is
small, but the field might change rapidly in configuration space, so
higher momentum components might be important in Fourier space.

In leading order of the weak field expansion, the FP equation is quadratic in
the fluctuations. Using translation symmetry we can write
\begin{equation}
{\cal A}^{FP}(\R) = {1 \over 2}\sum_{n_B,r_B} 
\rho(r_B) {\boldsymbol \chi}_{n_B}{\boldsymbol \chi}_{n_B+r_B} + O(\chi^4)\, ,
\label{fpsaquad}
\end{equation}
where the unknown couplings, to be determined by the FP equation, are denoted
by $\rho(r_B)$. Expanding also the blocking kernel in eq.~(\ref{fps1}), we get
at the quadratic level
\begin{eqnarray}
& &\frac{1}{2} \sum_{n_B,r_B} \rho(r_B){\boldsymbol 
\chi}_{n_B}{\boldsymbol \chi}_{n_B+r_B}= \\
& &\min_{\{{\pi}\}}\bigl[ 
\frac{1}{2} \sum_{n,r} \rho(r){\boldsymbol \pi}_{n}{\boldsymbol \pi}_{n+r}
+ 2\,\kappa_\sigma \sum_{n_B}\,({\boldsymbol \chi}_{n_B} - {1 \over 4}\sum_{n \in n_B}
{\boldsymbol \pi}_n )^2 \bigr]\, .\nonumber
\label{fpsequad}
\end{eqnarray}
Using the trivial relation between Gaussian integrals and minimization we
observe that eq.~(\ref{fpsequad}) is just the FP equation of a free scalar
theory:
\begin{eqnarray}
& &c \exp \bigl(-{1 \over 2}\sum_{n_B,r_B}
\rho(r_B) {\boldsymbol \chi}_{n_B}{\boldsymbol \chi}_{n_B+r_B}\bigr) = \\
& &\prod_{n}\int d{\boldsymbol \pi}_n \exp \bigl[
- \frac{1}{2} \sum_{n,r} \rho(r){\boldsymbol \pi}_{n}{\boldsymbol \pi}_{n+r}
- 2\,\kappa_\sigma \sum_{n_B}\,({\boldsymbol \chi}_{n_B} - {1 \over 4}\sum_{n \in n_B}
{\boldsymbol \pi}_n )^2 \bigr]\, .\nonumber
\label{freesc}
\end{eqnarray}
The same steps can be used for the Yang-Mills or QCD FP actions. In 
quadratic order the Yang-Mills FP equation, eq.~(\ref{fpyme}), goes over to the
FP problem of a free (Abelian) gauge theory, while the fermion part of the QCD
FP action, eq.~(\ref{fph}), becomes equivalent to the FP equation of free
Dirac fermions. In the next section we give a general solution of the free
field FP problem.

\section{The free field FP problem}

As we discussed in Sect.~7, in leading order of the weak field expansion the
FP equations of AF theories are reduced to that of free fields. We shall study
the free field problem in this section.

In all our previous considerations we applied lattice regularization.  For
free field theories without gauge symmetry we might also use a simple momentum
space cut-off, which leads quickly to some results we already referred to
in the previous sections\cite{wil,others,sh}. 
We present these results first, and then turn to the
problem of FP actions on the lattice.

\subsection{Scalar field with momentum space cut-off}

We work in Fourier space and constrain all the momenta to be smaller than
$\Lambda^{cut}$. As always, we use dimensionless quantities, the dimensions
are carried by the cut-off.

\noindent We write the quadratic action in the general form
\begin{equation}
{\cal A}(\phi) = {1 \over 2} \int_{|q|\le 1} \phi(-q) \,\rho(q)\, \phi(q)\, ,
\label{actqfree}
\end{equation}
and perform a RGT by integrating out the fields in the momentum range
$1/2<|q|\le 1$. We write
\begin{equation}
\phi(q) = \phi_0(q)+ \phi_1(q)\, ,
\label{fi01}
\end{equation}
where $\phi_0(q)$ and $\phi_1(q)$ are non-zero only for $|q|\le 1/2$ and
$1/2<|q| \le 1$, respectively. In the RGT we have to integrate out $\phi_1(q)$,
i.e. perform the path integral
\begin{equation}
\exp \bigl( - \tilde{{\cal A}}(\phi_0)\bigr) =
\prod_{{1 \over 2}<|q|\le 1} \int d\phi_1(q)\; \exp \bigl({\cal A}(\phi_0 + 
\phi_1)\bigr) \, .
\label{rgtqfree}
\end{equation}
Using the quadratic action in eq.~(\ref{actqfree}) we get
\begin{equation}
{\cal A}(\phi_0 + \phi_1) = {1 \over 2} \int_{|q|\le {1 \over 2}} \phi(-q) 
\rho(q) \phi(q)
+ {1 \over 2} \int_{{1 \over 2}<|q|\le 1} \phi(-q) \rho(q) \phi(q)\,.
\label{nocross}
\end{equation}
Writing eq.~(\ref{nocross}) into eq.~(\ref{actqfree}), the $\phi_0$-dependent
part can be brought out of the integral, while the $\phi_1$-integral gives a
field independent, irrelevant constant. By rescaling and relabeling the
$\phi_0$ field as
\begin{equation}
\phi'(q') = \zeta^{-1} \phi_0(q)\, \qquad q'=2\,q\,,\qquad 0\le |q'| \le 1 \, ,
\label{resrel}
\end{equation}
we get
\begin{multline}
\tilde{\A}(\phi_0) =\zeta^2 {1 \over 2} 
\int_{|q|\le {1 \over 2}} \phi'(-2\,q)\, \rho(q)\, \phi'(2\,q) = \\
\zeta^2 2^{-d}{1 \over 2}
\int_{|q|\le 1} \phi'(-q)\, \rho(q/2)\, \phi'(q)= \A'(\phi') \, .
\label{aza}
\end{multline}
The role of the rescaling factor $\zeta$ in eq.~(\ref{resrel}) is the same as
that of $b$ in eqs.~(\ref{avr},\ref{rgeq}),\break whose significance we shall
see in a moment. As eq.~(\ref{actqfree})
and eq.~(\ref{aza}) show, under the RGT
$\rho$ transforms as
\begin{equation}
\rho(q) \rightarrow \zeta^2 2^{-d} \rho(q/2) \, .
\label{rhotr}
\end{equation} 
Expand $\rho$ in powers of momenta 
\begin{equation}
\rho(q) = \alpha_0 + \alpha_2\,\sum_{\mu}\,q_\mu q_\mu +
\alpha_4\,(\sum_{\mu}\,q_\mu q_\mu)^2 +
\alpha_6\,(\sum_{\mu}\,q_\mu q_\mu)^3 \ldots \, .
\label{rhoexp}
\end{equation} 
For $\alpha_0$ small, the pole of the propagator $1/\rho(q)$ is close to $q=0$ and
$\alpha_0=m^2$, where $m$ is the mass of the free particle. We shall require that
the action density before and after the transformation goes over into the form of
classical field theory for small $q$-values, i.e. we fix the coefficient
of $\sum_{\mu}\,q_\mu q_\mu$ in $\rho$ to be 1. This condition fixes $\zeta$
\begin{equation}
\zeta^2 2^{-d} \sum_{\mu}\,{q_\mu \over 2} {q_\mu \over 2} = 
\sum_{\mu}\,q_\mu q_\mu \rightarrow \zeta=2^{{d+2 \over 2}}\,.
\label{zeta}
\end{equation} 
Eq.~(\ref{rhotr}) gives then the transformation rules for the other couplings
\begin{equation}
\alpha_0 \rightarrow 2^2\,\alpha_0 \, , \qquad 
\alpha_4 \rightarrow 2^{-2}\,\alpha_4\, ,
\qquad  \alpha_6 \rightarrow 2^{-3}\,\alpha_6\, , \ldots \; .
\label{coupltr}
\end{equation}
Eq.~(\ref{coupltr}) shows that the operator $\phi^2$ (the coupling $\alpha_0$
or $m^2$) is relevant with an eigenvalue $\lambda_0=4$, all the other operators
(couplings) which contain four or more derivatives are irrelevant. The
largest irrelevant eigenvalue is 1/4.

Replacing the region of allowed momenta of a cut-off sphere by a cut-off
hypercube, the regularization will violate rotation symmetry. Only hypercubic
symmetry remains, like on the lattice. In this case we have to allow in
eq.~(\ref{rhoexp}) terms like $\beta_4 \sum_\mu (q_\mu q_\mu q_\mu q_\mu)$, which
is not $O(d)$ invariant, but hypercubic symmetric. The coupling $\beta_4$ is
also irrelevant, $\beta_4 \rightarrow 1/4  \beta_4$ under the RGT. That means that
rotational symmetry is restored as the RGT is iterated, as we go towards
the infrared.

If we start with $\alpha_0 =0$, we are and remain on the critical surface and
run rapidly towards the FP which, in this case, has the simple form
\begin{equation}
{\cal A}^{FP}(\phi) = {1 \over 2} \int_{|q|\le 1} \phi(-q) q^2 \phi(q) \, ,
\label{fpfrecq}
\end{equation}
describing a free massless scalar particle. As we mentioned before, the form
of the FP depends on the block transformation (typically, it will look more
complicated on the lattice), but the conclusions concerning the relevant,
marginal and irrelevant eigenvalues is independent of that.

Consider now small perturbations around the FP, eq.~(\ref{fpfrecq}),
like $\lambda_4 \phi^4,\,\lambda_6 \phi^6,\cdots$, with $\lambda_4,\,\lambda_6,
\cdots$ small, as
discussed in Sect.~3.4
\begin{multline}
\A(\phi) = \A^{FP}(\phi) +
{\alpha_0 \over 2} \int_q \phi(-q) \phi(q) + 
\lambda_4 \int_{q_1,\ldots,q_4} \phi(q_1)\ldots \phi(q_4)
\delta(q_1+\ldots +q_4) +\\
\lambda_6 \int_{q_1,\ldots,q_6} \phi(q_1)\ldots\phi(q_6)
\delta(q_1+\ldots +q_6) +\ldots \, ,
\label{fppertq}
\end{multline}
where only three of the perturbations are written out explicitly. Using
this action ${\cal A}$ in eq.~(\ref{rgtqfree}), expanding the exponent in
$\alpha_0, \lambda_4, \lambda_6 \ldots$ up to linear order, performing the
Gaussian $\phi_0$-integrals and rescaling and relabeling as in
eq.~(\ref{resrel}), we get

\begin{eqnarray}
\alpha_0&\rightarrow&(\alpha_0 + 6\,c \lambda_4  + 
15 \,c^2 \lambda_6  + \ldots)\zeta^2 2^{-d} \, , \nonumber\\
\lambda_4&\rightarrow&(\lambda_4  + 
15 \,c^2 \lambda_6 + \ldots)\zeta^4 2^{-3d} \, ,\label{linrgtq}\\
\lambda_6&\rightarrow&(\lambda_6  + 
 \ldots)\zeta^6 2^{-5d} \, , \nonumber
\end{eqnarray}
with
\begin{equation}
c = \int^{|q|=1}_{|q|={1 \over 2}} {1 \over q^2}\, .
\label{cint}
\end{equation}
The corresponding $T$-matrix (eq.~(\ref{tmatrix})) is a triangular matrix, the
eigenvalues $\lambda$ are just the diagonal elements
\begin{eqnarray}
\lambda_1=2^2\,,&&\qquad h^{(1)}=\int_{x} \phi^2(x)\, , \nonumber\\
\lambda_2=2^{4-d}\,,&&\qquad h^{(2)}=\int_{x} 
\bigl[ \phi^4(x)+ \alpha\phi^2(x)\bigr] \, ,\label{lamh}\\
\lambda_3=2^{6-2d}\,,&&\qquad h^{(3)}=\int_{x} \bigl[\phi^6(x)+\beta_1\phi^4(x)
+\beta_2\phi^2(x)\bigr] \,. \nonumber
\end{eqnarray}
As eq.~(\ref{lamh}) shows, there is a simple relation between the eigenvalue
$\lambda_\theta$ and the dimension $d_\theta$ of the operator $\theta(x)$ which
has the highest dimension in the corresponding eigenoperator $h$:
\begin{equation}
\lambda_\theta = 2^{d-d_\theta}
\label{lamtheta}
\end{equation}
In the eigenvector $h$ the highest dimensional operator $\theta(x)$ is mixed
with lower dimensional operators.

In $d=4$, operators with dimension less than, equal, larger than 4 are
relevant, marginal and irrelevant, respectively. This result is also valid in 
gauge and fermion theories. In $SU(N)$ Yang-Mills theory the lowest
dimensional gauge invariant operator is Tr$(F_{\mu \nu}F_{\mu \nu})$, which is
marginal.

\subsection{Lattice FP actions in different free field theories}
Consider first a real scalar field. Denote the fields on the coarse and on the
fine lattice by $\chi$ and $\phi$, respectively. The lattice units are $a_0$
and $a_1=a_0/2$. Assume, we are in the FP, and perform a RGT:
\begin{multline}
c \exp \bigl(-{1 \over 2}\sum_{n_B,r_B}
\rho(r_B) \chi_{n_B} \chi_{n_B+r_B}\bigr) = \\
\prod_{n}\int d \phi_n \exp \bigl[
- \frac{1}{2} \sum_{n,r} \rho(r) \phi_{n} \phi_{n+r}
- {1 \over 2}\kappa \sum_{n_B}\,( \chi_{n_B} - 
b \sum_{n}\omega(2n_B-n) \phi_n)^2 \bigr] \, .
\label{scalar82}
\end{multline}
Multiply this equation by $\chi_{n'_B}\chi_{n''_B}$ and integrate over the
$\chi$ fields. Using eq.~(\ref{zpez}) we obtain
\begin{equation}
\langle \chi_{n'_B}\chi_{n''_B} \rangle = b^2 \sum_{n',n''}
\omega(2n'_B-n')\omega(2n''_B-n'')\langle \phi_{n'}\phi_{n''} \rangle
+ {1 \over \kappa} \delta_{n'_B,n''_B} \, .
\label{step1s}
\end{equation}
In Fourier space eq.~(\ref{step1s}) has the form
\begin{multline}
\frac{1}{\tilde {\rho}(q_B)} = b^2 {1 \over 2^d} \sum_{l}
\tilde {\omega}({q_B \over 2}+\pi l) \frac{1}{\tilde {\rho}({q_B \over 2}+\pi l)}
\tilde {\omega}^*({q_B \over 2}+\pi l)
+ {1 \over \kappa} ,\\ 
l=(l_1,\ldots,l_d)\,, l_\mu=0,1 \, .
\label{step1sq}
\end{multline}
The Fourier transformation is defined as
\begin{equation}
f(n) = \int^{\pi}_{-\pi} \ldots \int^{\pi}_{-\pi}{d^dq \over (2\pi)^d}
\exp(iqn) \tilde {f}(q) \, .
\label{foutr}
\end{equation}

We demand, as discussed in Sect.~8.1, that any action should go over the
classical continuum action for smooth fields. That requires that $\rho(q_B)
\rightarrow q_B^2 +O(q_B^4)$ for small $q_B$. (The FP action is on the critical
surface, the mass parameter $\alpha_0$ in eq.~(\ref{rhoexp}) is zero.)
Considering eq.~(\ref{step1sq}) in the $q_B \rightarrow 0$ limit, the $l\not=
0$ contribution on the r.h.s. can be neglected and we obtain 
\begin{equation}
{1 \over q_B^2} = b^2 {1 \over 2^d}|\tilde {\omega}(0)|^2 
{1 \over (q_B/2)^2} \, .
\label{residue}
\end{equation}
The normalization condition, eq.~(\ref{omnorm}), gives $\tilde{\omega}(0) =1$
and so
\begin{equation}
b= 2^{{d-2 \over 2}} \, .
\label{bpar}
\end{equation}

We shall use the relation of eq.~(\ref{step1s}) recursively, i.e. we connect the
propagator $\langle \phi_{n'}\phi_{n''} \rangle$ on the r.h.s. of
eq.~(\ref{step1s}) with the propagator on the lattice with unit
$a_2=a_0/4$. Denote the field on this lattice by $\eta$ and the lattice points
by $n_f$:
\begin{multline}
\langle \chi_{n'_B}\chi_{n''_B} \rangle = b^4 
\sum_{n',n'_f} \omega(2n'_B-n')\omega(2n'-n'_f)\\
\times \sum_{n'',n''_f} \omega(2n''_B-n'')\omega(2n''-n''_f)
\langle \eta_{n'_f}\eta_{n''_f} \rangle + \\
{1 \over \kappa} \bigl[ \delta_{n'_B,n''_B} +
b^2 \sum_{n} \omega(2n'_B-n) \omega(2n''_B-n) \bigr] \, .
\label{step2s}
\end{multline}
Iterating this equation further, after $j$ steps one arrives at a lattice with
lattice unit $a_j=a_0/2^j$. For $j \rightarrow \infty$, the lattice will be
infinitely fine and the propagator for this lattice can be replaced by the
propagator in the continuum. This way, the FP propagator
$\langle\chi_{n'_B}\chi_{n''_B} \rangle$ will be expressed explicitly in
terms of the propagator in the continuum and of the product and sum of the
averaging function $\omega$. The inverse propagator gives the FP action.

If the block transformation is non-overlapping (see the paragraph after
eq.~(\ref{normc})), the sum in the last term of eq.~(\ref{step2s}) is $ \sim
1/\kappa \delta_{n'_B,n''_B}$. This remains true in every order of the
iteration. Although for overlapping transformations this is not true anymore,
the $\sim 1/\kappa$ contribution to the propagator remains local. The
non-locality of the propagator is carried completely by the first term in
eq.~(\ref{step2s}). 

Without going through the detailed derivation, we quote the final results
only.
Let us start with an example: take the simple non-overlapping transformation,
where $\omega(2n_B-n)=2^{-d}$ if $n$ is in the hypercube  whose center is
indexed by $n_B$ and zero otherwise (Sect.~3.1). The FP propagator in Fourier
space has the form\cite{bell,iwa,gaku}
\begin{equation}
\frac{1}{\tilde {\rho}(q)} = \sum_{l \in Z^d} {1 \over (q+ 2\pi l)^2}
\prod_{\mu}\frac{\sin^2({q_\mu \over 2}+\pi l_\mu)}{({q_\mu \over 2}+\pi
l_\mu)^2}
+{4 \over 3\kappa} \, ,
\label{fpexa}
\end{equation}
where the summation is over integer vectors $l\!=\!(l_1,\cdots\!,l_d)$ and
$(q\!+\!2\pi l)^2\!=\!\sum_{\mu}(q_\mu\! +\! 2\pi l_\mu)^2$.
The $\sim 1/\kappa$ term in eq.~(\ref{fpexa}) is a constant in Fourier
space, hence $\sim \delta_{n,n'}$ in configuration space, which is the
general result for non-overlapping transformations. The second factor under
the sum in eq.~(\ref{fpexa}) is regular, it is the Fourier transform of the
product of the local averaging function $\omega$. The pole singularities come
entirely from the $(q+2\pi l)^{-2}$ piece. These poles define the spectrum of
the FP theory on the lattice. Fix the spatial part of the momentum to be 
${\bf q} ={\bf p}, p_i \in (-\pi, \pi)$, and find the pole singularities in 
the complex $q_d$-plane. There is an infinite sequence of poles on the
imaginary axis
\begin{equation}
q_d = \pm i\, |{\bf p}+2\pi {\bf l}|\,, \qquad {\bf l} \in Z^{d-1} \, .
\label{poles}
\end{equation} 
Denoting ${\bf p}+2\pi {\bf l}$ by ${\bf k}, k_i \in (-\infty, \infty)$, we
get the exact relativistic dispersion relation of a massless particle
\begin{equation}
-i\,q_d = E({\bf k}) = |{\bf k}| \, .
\label{energies}
\end{equation}
Through the $l$-summation in eq.~(\ref{fpexa}), the full exact continuum
spectrum is reproduced (the momentum ${\bf k}$ is not constrained to the
Brillouin zone), even though we are on the lattice, which resolves the large $k$
waves very poorly. This result is to be compared with the dispersion relation
of the standard nearest-neighbour action (eq.~(\ref{latlagr}) with $V=0$). If
$k$ is close to the end of the Brillouin zone, the dispersion relation is wrong
by a factor of 2 or more.

Performing the sum in eq.~(\ref{fpexa}) and calculating $\rho(r)$ by Fourier
transformation is an easy numerical exercise. The function $\rho(r)$ and hence
the FP action is local, as expected: it decays exponentially $\sim
\exp(-\gamma |r|)$, as discussed in Sect.~2.4. The size of $\gamma$ which
defines how short ranged the action is, depends on the free parameter
$\kappa$. The value $\kappa \approx 4$ is optimal. For $d=2$, we have $\gamma
\approx 3.4$ for $\kappa=4$. In this case $\rho(r)$ is strongly dominated by
the nearest neighbour and diagonal couplings 
as table~1 shows\cite{hn}.

\begin{table}  
\caption{The couplings of the free field FP action 
at a distance $r=(r_0,r_1)$ for the optimal choice of the
block transformation with $\kappa\!=\!4$.
Note that in our convention, for the standard action 
the  only non--vanishing entry in this list would be
$\rho_{ST}(1,0)=-1$.}\vskip3mm
\begin{center}
\begin{tabular}{|c|c||c|c|}
\hline
$r$ & $\rho(r)$ & $r$ & $\rho(r)$ \\
\hline
(1,0) & $-0.61802 $           & (4,0) & $-2.632\cdot 10^{-6}$ \\
(1,1) & $-0.19033 $           & (4,1) & $ 7.064\cdot 10^{-7}$ \\
(2,0) & $-1.998\cdot 10^{-3}$ & (4,2) & $ 1.327\cdot 10^{-6}$ \\
(2,1) & $-6.793\cdot 10^{-4}$ & (4,3) & $-7.953\cdot 10^{-7}$ \\
(2,2) & $ 1.625\cdot 10^{-3}$ & (4,4) & $ 6.895\cdot 10^{-8}$ \\
(3,0) & $-1.173\cdot 10^{-4}$ & (5,0) & $-8.831\cdot 10^{-8}$ \\
(3,1) & $ 1.942\cdot 10^{-5}$ & (5,1) & $ 3.457\cdot 10^{-8}$ \\
(3,2) & $ 5.232\cdot 10^{-5}$ & (5,2) & $ 3.491\cdot 10^{-8}$ \\
(3,3) & $-1.226\cdot 10^{-5}$ & (5,3) & $-3.349\cdot 10^{-8}$ \\
      &                       & (5,4) & $ 8.408\cdot 10^{-9}$ \\
      &                       & (5,5) & $-1.657\cdot 10^{-10}$ \\
\hline
\end{tabular}
\end{center}
\end{table}

Actually, $\rho(r)$ is a perfect discretization of the continuum
Laplacian. The question raised by this observation is, whether the FP idea can
be used in solving partial differential equations, like the Navier-Stokes
equation, numerically\cite{wie2} .

Let us present now the general result. The fields can be scalars, fermions or
vectors. For the blocking kernel we write
\begin{equation}
\kappa \sum_{n_B}
\bigl( \bar{\chi}^\alpha_{n_B}-b\sum_{n,\beta}\bar{\phi}^\beta_n
\omega^\dagger_{\beta,\alpha}(2n_B-n) \bigr)
\bigl( \chi^\alpha_{n_B}-b\sum_{n,\beta}\omega_{\alpha,\beta}(2n_B-n)
\phi^\beta_n \bigr) \, ,
\label{kernelx}
\end{equation}
where the bar above the fields denotes complex or Dirac 
conjugation\footnote{For a real scalar or vector field $\bar{\phi}=\phi$ and 
we include a factor of 1/2 both in the action and in the kernel, as in
eq.~(\ref{scalar82}).}, the index $\alpha$ can be a Lorentz or Dirac 
index\footnote{ We shall
consider the block transformation only where $\omega$ is diagonal in the Dirac
indices.}. The FP propagator is given by
\begin{equation}
\tilde{D}^{FP}(q) = \lim_{n \to \infty} \tilde{D}^{(n)}(q) \, ,
\end{equation}
where
\begin{eqnarray}
\tilde{D}^{(n)}(q)& =&(2^{-d}b^2)^n \sum^{2^n-1}_{l_\mu=0}
\tilde{\Omega}^{(n)}(\frac{q+2\pi l}{2^n}) 
\tilde{D}^{(0)}(\frac{q+2\pi l}{2^n})
\tilde{\Omega}^{\dagger (n)}(\frac{q+2\pi l}{2^n}) \\
& & + {1 \over \kappa} \sum^{n-1}_{j=0} \tilde{Q}^{(j)}(q) \, ,\nonumber
\label{recgen}
\end{eqnarray}
where $\Omega$ and $Q$ are defined as
\begin{eqnarray}
\tilde{\Omega}^{(j)}&=& \tilde{\omega}(2^{-1})\tilde{\omega}(2^{-2})
\ldots \tilde{\omega}(2^{-j}) \, ,
\label{bigom}\\
\tilde{Q}^{(j)}(q) &=& (2^{-d}b^2)^j \sum^{2^j-1}_{l_\mu=0}
\tilde{\Omega}^{(j)}(\frac{q+2\pi l}{2^j})
\tilde{\Omega}^{\dagger (j)}(\frac{q+2\pi l}{2^j})\, .
\label{bigq}
\end{eqnarray}
The rescaling factor $b$ is given by
\begin{equation}
b=2^{d_\phi} \, ,
\label{bgen}
\end{equation}
where $d_\phi$ is the dimension of the field $\phi$. For non-overlapping
transformations $\tilde{Q}^{(j)}(q)$ is independent of $q$, in the overlapping
case it is an analytic function of $q$.
For illustration, consider free Dirac fermions in $d=4$ using the overlapping
block transformation defined by the eqs.~(\ref{omc},\ref{cflat}). 
In this case\cite{ku}
\begin{eqnarray}
\tilde{\omega}(q)&=&\prod_{\mu} \cos^2({q_\mu \over 2})\, , \nonumber\\
\tilde{\Omega}^{(j)}(q)&=&\prod_{\mu} 2^{-2j}\frac{\sin^2({q_\mu \over 2})}
{\sin^2({q_\mu \over 2^{j+1}})}\, ,
\label{exa2}\\
\tilde{Q}^{(j)}(q) &=&{2^{-j} \over 81}\prod_{\mu}\bigl[
(1+4^{-j})\cos(q_\mu) + (2+4^{-j}) \bigr] \, . \nonumber
\end{eqnarray}
For the initial fermion propagator $D^{(0)}$ one might take the Wilson fermion
propagator (Sect.~5.3). Eq.~(\ref{recgen}) and eq.~(\ref{exa2}) gives the FP
fermion propagator
\begin{equation}
D^{FP}(q) = -i\alpha^\mu(q) \gamma_\mu + \beta(q) \, ,
\label{flatfp}
\end{equation}
where
\begin{equation}
\alpha^\mu(q) = \sum_{l \in Z^d} \frac{q_\mu + 2\pi l_\mu}
{(q+2\pi l)^2} \prod_{\nu}\frac{\sin^4({q_\nu \over 2})}
{({q_\nu \over 2}+\pi l_\nu)^4} \, 
\label{flatfa}
\end{equation}
and
\begin{eqnarray}
\beta(q)& =& {1 \over 81 \kappa} \bigl[ c_1\prod_{\mu}\cos(q_\mu)+
c_2\bigl(\cos(q_1)\cos(q_2)\cos(q_3)+ \ldots \\
& & + (\cos(q_2)\cos(q_3)\cos(q_4)\bigr)+
c_3\bigl(\cos(q_1)\cos(q_2)+ \ldots +\cos(q_3)\cos(q_4)\bigr)+\nonumber \\
& & c_4 \sum_{\mu} \cos(q_\mu) + c_5 \bigr] \, .\nonumber
\label{flatfb}
\end{eqnarray}
The constants $c_1,\dots,c_5$ result from different geometric sums and have
the value $c_1=0.592581,~c_2=1.388406,~c_3=3.349502,~c_4=8.336553,~
c_5=102.410571$. This FP propagator has all the features we expected. The
spectrum is given by the $\sim \alpha^\mu \gamma_\mu$ part and it is
exact. The $\sim 1/\kappa$ part is analytic in $q$-space and ultralocal
(Sect.~2.4) in configuration space. The propagator is not chiral symmetric.
The chiral symmetry breaking term is $\sim 1/\kappa$ comes entirely from the
non-chiral invariant block transformation.

\section{Instanton solutions and the topological charge on the lattice}  

In this section we shall consider the $O(3)$ non-linear $\sigma$-model in
$d=2$. The arguments and results can be immediately generalized to Yang-Mills
theories in $d=4$.

As we discussed in Sect.~5.1, the configurations of the $O(3)$ model fall into
different topological sectors characterized by an integer number $Q$, the
topological charge of the configuration. While configurations from the same
topological sector can be continuously deformed into each other, this is not
possible for configurations with a different charge $Q$. The charge $Q$ is the
number of times the internal sphere is wrapped as the coordinate sphere is
traversed. It may be defined as
\begin{equation}
Q={1 \over 8\pi} \int d^2x \epsilon_{\mu \nu}\Sv (\partial_\mu \Sv \times 
\partial_\nu \Sv) \, ,
\label{topchscont}
\end{equation}
and it is related to the action by the inequality
\begin{equation}
\label{ineq}
{\cal A} \geq 4\,\pi |Q|\, .
\end{equation} 
If, for a given configuration the equality is satisfied, the configuration
minimizes the action for the given topological charge $Q$ and is therefore a
solution of the equations of motion. In order to show the inequality in
eq.~(\ref{ineq}) one writes down the identity
\begin{equation}
{1 \over 4} \int [\partial_\mu \Sv + 
\epsilon_{\mu \nu} (\Sv \times 
\partial_\nu \Sv)]^2 = 
{1 \over 2} \int (\partial_\mu \Sv)^2 - {1 \over 2} \int
\epsilon_{\mu \nu}\Sv (\partial_\mu \Sv \times \partial_\nu \Sv)
 \, ,
\label{ident}
\end{equation}
which follows easily from vector product identities and from $\partial_\mu
\Sv^2= 0$. The l.h.s. of eq.~(\ref{ident}) is non-negative, while on the
r.h.s. a combination of the action and of the topological charge enters
leading to eq.~(\ref{ineq}).

\subsection{Instanton solutions of the FP action}

Replacing the continuous Euclidean space by a discrete set of points, the
notion of a `continuous deformation of the configuration' is lost and the
definition of the topological charge becomes problematic. A possibility is
('field theoretical definition') to discretize eq.~(\ref{topchscont}) in some
simple way taking care of the surviving discrete symmetries \cite{Vecc}. The
nice geometrical properties of the topological charge are lost this way, it is
not an integer valued functional of the fields any more and it suffers from a
multiplicative renormalization on the lattice \cite{DiG}. 
Another possibility is the
'geometrical definition' \cite{geo} which assigns an integer charge to each
configuration. However, this charge receives, in general,  unphysical 
contributions from topological artifacts as discussed below.

An additional problem is that
the lattice introduces a scale (lattice unit $a$), classical scale invariance
is broken and no scale invariant (instanton) solutions are expected to exist.
Of coarse, if the radius $\rho$ of the instanton is much larger than the
lattice unit $a$, the values of the continuum solution in the lattice points
define a configuration which is `almost' a solution of the lattice
Euler-Lagrange equations. The action of this quasi-solution will be
\begin{equation}
{\cal A} = 4\,\pi |Q|\,\bigl( 1 + c\,{a^2 \over \rho^2} + 
O({a^4 \over \rho^4}) \bigr) \, .
\end{equation}
The size and sign of $c$ depends on the details of the form of the lattice
action. Depending on the sign of $c$, large instantons shrink or grow to get
a smaller action.

An even more serious problem is the possible existence of dislocations,
topological artifacts. These are small, $O(a)$ objects which, especially when
using the geometrical definition of the topological charge,
might contribute to the
topological charge and have an action below  $ 4\,\pi |Q|$. The danger is that
these low-action, non-physical objects dominate certain topology related
expectation values (like $<Q^2>/V$, where $V$ is the volume of the system) and
might even spoil the continuum limit completely \cite{LuWo}.

There are many high energy physicists who believe that fluctuating instanton
configurations play an important role in QCD, especially for questions where
chiral symmetry is relevant. If a discretized action describes the classical
topological properties of the theory poorly, then it might happen that
continuum quantum physics will be reproduced on very fine lattices only. This
would increase the numerical difficulties which are not small anyhow.

The FP actions offer a solution to this 
problem \cite{hn,bb1,n7,n8,n22,n23,n29}.
We show now that the Euler-Lagrange equations of the FP action have scale
invariant instanton solutions. We demonstrate first the following 
statement \cite{hn}:

If the configuration $\R$ on the coarse lattice satisfies the Euler-Lagrange
equations of the FP action
and it is a local minimum of ${\cal A}^{FP}(\R)$, then the configuration
$\Sv_{min}(\R)$ on the finer lattice which minimizes the r.h.s. of the FP
equation eq.~(\ref{fps1}) satisfies the Euler-Lagrange equations of motion
as well. In addition, the
value of the action remains unchanged, ${\cal A}^{FP}(\Sv_{min})= {\cal A}^{FP}(\R)$.

This statement is easy to show. Since $\R$ is a solution of the classical
equations of motion $\delta{\cal A}^{FP}/\delta\R=0$, then the configuration
$\Sv=\Sv_{min}(\R)$ should satisfy the equation
\begin{equation}
\Si_{n_B} = \lambda_{n_B} \R_{n_B} \, ,
\label{s9}
\end{equation}
for any $n_B$. Here $\Si_{n_B}$ is the sum of the four
$\Sv_n$ spins in the block, eq.~(\ref{sigavr1}). Eq.~(\ref{s9}) follows by
taking the variation $\delta/\delta\R$ of the FP equation,
eq.~(\ref{fps1}). Since we assumed that the configuration $\R$ is a local
minimum of ${\cal A}^{FP}$, we have $\lambda_{n_B}\ge 0$. On the other hand, the
$\sim \kappa_\sigma$ term in eq.~(\ref{fps1})
\begin{equation}
-\kappa_\sigma \sum_{n_B}\bigl( \R_{n_B}\Si_{n_B} - |\Si_{n_B}| \bigr) \, ,
\label{s10}
\end{equation}
takes its absolute minimum (zero) on the configuration $\Sv_{min}(\R)$
satisfying  eq.~(\ref{s9}). Since $\Sv_{min}(\R)$ is the minimum of the sum of
eq.~(\ref{s10}) and ${\cal A}^{FP}(\Sv)$, it follows that  $\Sv_{min}(\R)$ is a
stationary point of ${\cal A}^{FP}$:
\begin{equation}
\left(\frac{\delta{\cal A}^{FP}(\Sv)}{\delta \Sv}\right)_{|\Sv=\Sv_{min}} = 0 \, .
\end{equation}
Since eq.~(\ref{s10}) is zero if eq.~(\ref{s9}) is satisfied, we get
${\cal A}^{FP}(\R)={\cal A}^{FP}(\Sv_{min})$, as we wanted to show.

According to this statement, if ${\cal A}^{FP}$ has an instanton solution of size
$\rho$, then there exist instanton solutions of size $2\rho,
2^2\rho,\ldots$ with the value of the action being exactly $4\pi$ for all
these instantons (scale invariance). The value $4\pi$ follows from the fact
that very large instantons are smooth on the lattice and then any valid
lattice action gives the continuum value.

\begin{figure}[htb]
\begin{center}
\leavevmode
\epsfxsize=110mm
\epsfbox{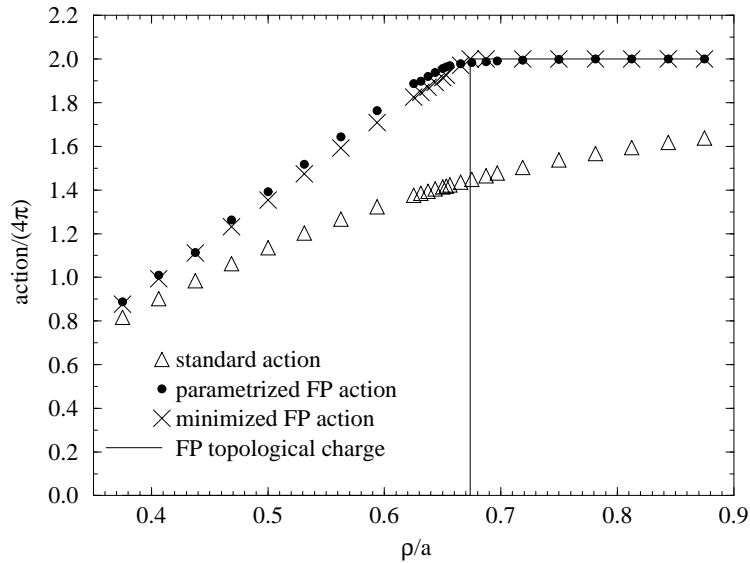}
\end{center}
\caption{Actions and charge of instantons with radii of the order of one
lattice spacing
}
\label{fig4}
\end{figure}

It is important to observe that the reverse of the statement is not true: if
the configuration $\Sv$ is a solution, then the configuration $\R$, where
$\R_{n_B}= \Si_{n_B}/|\Si_{n_B}|$, ($\Si_{n_B}$ is defined in eq.~(\ref{sigavr1}))
is not
necessarily a solution. The proof fails because for this configuration $\R$
the minimizing configuration is not necessarily equal to $\Sv$ itself. Really,
consider eq.~(\ref{fps1}) with this $\R$ configuration. We have to find
$\Sv_{min}$ minimizing the r.h.s.. It is easy to see that $\Sv$ itself (which
served to produce the configuration $\R$) is a minimum. It is, however, not
necessarily the absolute minimum. One can study the problem in every detail
numerically \cite{n7,n22}. It turns out that if the configuration $\Sv$ has an
instanton with $\rho \gg a$, then for the blocked $\R$ in eq.~(\ref{fps1})
$\Sv$ itself is the absolute minimum. In this case, $\R$ is also a solution of
the Euler-Lagrange equations of motion and describes an instanton with radius
$\rho/2$. If, however, $\rho/2$ is smaller than $a$ (actually, smaller than $\sim
0.7 a$ for the FP defined by the block transformation eq.~(\ref{fps1})) a
competing, distinctly different configuration becomes the absolute
minimum. The configuration $\R$ will not be a solution of the classical
equations of motion, it will not describe an instanton anymore. The instantons
with radius smaller than $\sim 0.7 a$  'fall through the lattice', they do not exist.
The value of the action is exactly $2\pi$ until $\rho$ is larger than $\sim
0.7a$. At the moment the instanton 'falls through the lattice', a new minimum takes
over, a non-analyticity (a knick) occurs, and the value of the action begins 
to decrease. At the same moment, the topological charge drops to
zero. This non-analyticity has no direct relation to the locality properties
of the action. The action is local. 
Fig.~\ref{fig4} is the result of a detailed numerical study in the $O(3)$
non-linear $\sigma$-model \cite{n7}.
The existence of scale invariant instanton solutions, and the process of a
small instanton 'falling through' can be studied analytically on the example
of the quantum rotor \cite{n23}. The basic variables are angles in this case
and the non-analyticity discussed above comes through a mod($2\pi$)
prescription in the FP action. The FP action is local, in this case even
ultralocal (nearest neighbour) \cite{n23}.

We remark that in $SU(2)$ Yang-Mills theory the smallest instanton the lattice
can support has a size of $\sim 0.7 a$ quite similarly to the $O(3) \,
\sigma$-model case.

To prove the statement above we only had to show that the blocking kernel
$T^{(\infty)}_\sigma(\R,\Sv)$ in the FP equation, eq.~(\ref{fps}), is zero at
its absolute minimum in $\R$. This is true in all the cases we considered
($\sigma$-model, Yang-Mills theory) simply by construction: the normalization
condition of the kernel reads
\begin{equation}
\int D\R \exp \bigl(-\beta T_\sigma(\R,\Sv) \bigr)=1 \, ,
\end{equation}
which implies in the $\beta \rightarrow \infty$ limit that
$\min_{\R}T^{(\infty)}_\sigma(\R,\Sv) =0$. This can be seen explicitly in
eq.~(\ref{fps1}) and eq.~(\ref{sigt2}) for two different RGTs in the
$\sigma$-model, and in eqs.~(\ref{fpyme},\ref{fpymnorm}) in the Yang-Mills
theory.

The way this theorem works for the instanton solutions in the 
$O(3)\,  \sigma$-model and in the $SU(2)$ Yang-Mills 
theory has been tested in detail numerically.

\subsection{Fixed point operators}

Until now we have discussed the construction of the FP action only. We can add
to the action different operators $O_i$ multiplied by infinitesimal sources
$c_i$, and consider the RGT
\begin{eqnarray}
& & \exp\bigl[-\beta'{\cal A}'(\R)-c'_1\,O'_1(\R)-c'_2\,O'_2(\R)-\ldots\bigr]=\\
& & \int D\Sv \exp\bigl[-\beta\bigl({\cal A}(\Sv)+T_\sigma(\R,\Sv)\bigr)
-c_1\,O_1(\Sv)-c_2\,O_2(\Sv)-\ldots\bigr]\, .\nonumber
\label{rgtop}
\end{eqnarray}
Start with the FP action ${\cal A}^{FP}(\Sv)$ and take the $\beta \rightarrow
\infty$ limit. Using eq.~(\ref{fps}) we obtain
\begin{equation}
c'_1\,O'_1(\R)=c_1 O_1(\Sv_{min}(\R)),\qquad
c'_2\,O'_2(\R)=c_2 O_2(\Sv_{min}(\R)), \ldots \, .
\label{recop}
\end{equation} 
The FP form of the operator is reproduced by the RGT:
\begin{equation}
\lambda_1\,O^{FP}_1(\R)=O^{FP}_1(\Sv_{min}(\R))\,,\qquad
\lambda_2\,O^{FP}_2(\R)=O^{FP}_2(\Sv_{min}(\R))\,,\ldots\, , 
\label{fpop}
\end{equation}  
with $\lambda_i=c'_i/c_i$. The analysis above is similar to the general
discussion in Sect.~3.4, except that here we also allow operators
which one can not consider as part of the action (for example local operators
which are not summed over the lattice points).
The eigenvalues $\lambda$ are determined by simple dimensional analysis if the
FP is the Gaussian FP (which is the case for 
free fields and for AF theories), Sect.~8. 

Let us consider examples. Denote the field variables of a free scalar theory in $d$
dimension by $\phi_n$ and $\chi_{n_B}$ on the fine and coarse lattice,
respectively. Consider a RGT with scale factor 2. The FP field at the point $n_B$ is
constructed out of the $\chi$ fields in the local neighbourhood of $n_B$ and
has the transformation law
\begin{equation}
2^{-{d-2 \over 2}}\,\varphi^{FP}(\chi;n_B) = 
\varphi^{FP}(\phi_{min}(\chi);n=2n_B)\, ,
\label{fpscf}
\end{equation} 
where $\phi_{min}(\chi)$ is the minimizing configuration of the quadratic FP
equation of the free field theory.

As a second example, consider a current $J_\mu(n)$ in QCD. The FP current is a
local combination of the fermion fields connected by the product of gauge
matrices to assure gauge invariance. Beyond basic symmetries (which depend on
the type of current we consider) the details are determined by the
equation
\begin{equation}
2^{-3}\, J_\mu(\bar{\chi},\chi,V;n_B)=
J_\mu\bigl( \bar{\psi}_{st}(\bar{\chi},U_{min}),
\psi_{st}(\chi,U_{min}),U_{min}(V); n=2n_B\bigr)\, ,
\label{fpcurrent}
\end{equation} 
where $U_{min}$ is the minimizing configuration of the FP equation,
eq.~(\ref{fpyme}), while $\bar{\psi}_{st}, \psi_{st}$ are the configurations
which make the r.h.s. of eq.~(\ref{fph1}) stationary.

Only a few cases were treated explicitly until now: the FP free field
eq.~(\ref{fpscf}) and the FP Polyakov loop \cite{bb1}, further  
the FP topological charge
\cite{n7,n8,n23}, which will be discussed in the next subsection.

\subsection{The FP topological charge}

As we discussed before,
there are many ways to discretize the continuum definition of the topological
charge, eq.~(\ref{topchscont}). One might replace the derivatives by finite
differences taking care of the basic symmetries of the expression. This is the
so called `field theoretical' definition \cite{Vecc}. 
The corresponding topological charge
on the lattice will not be an integer number. This creates immediately a
complicated renormalization problem since there will be contributions to this
operator even in perturbation theory \cite{DiG}. 
An alternative possibility is to
construct an operator on the lattice which gives an integer number on any
configuration ('geometric definition') \cite{geo}
and so it is protected in perturbation
theory. This definition assigns, however, a non-zero topological charge to
different topological dislocations of size $O(a)$ which have nothing to do
with instantons and which heavily distort the results for quantities related to 
topology \cite{LuWo}. 

The FP topological charge used together with the FP action avoids these
difficulties.
Since $Q$ is a dimensionless quantity it satisfies the FP equation:
\begin{equation}
Q^{FP}(\R) = Q^{FP}(\Sv_{min}(\R))\, .
\label{topchsfp1}
\end{equation} 
One can solve this equation the following way. Let us denote the minimizing
configuration associated with $\R$ by $\Sv^{(1)}_{min}(\R)$ ($\equiv
\Sv_{min}(\R)$ in eq.~(\ref{topchsfp1})). Consider now the FP equation,
eq.~(\ref{fps}), with $\Sv^{(1)}_{min}$ on the l.h.s . The corresponding
minimizing configuration $\Sv^{(2)}_{min}=\Sv^{(2)}_{min}(\Sv^{(1)}_{min}(\R))$
lives on a lattice whose lattice unit is a factor of $2^2$ smaller than that
of the lattice of $\R$. Iterating  eq.~(\ref{topchsfp1}) this way, we get
\begin{equation}
Q^{FP}(\R) = Q^{FP}(\Sv^{(j)}_{min}(\R))\, ,
\label{topchsfpi}
\end{equation} 
where $\Sv^{(j)}_{min}(\R)$ lives on a very fine lattice (containing big
instantons with small fluctuations) if $j$ is large. On such configurations
any lattice definition $Q_{latt}$ gives the correct result
\begin{equation}
Q^{FP}(\R) = \lim_{j \to \infty} Q_{latt}(\Sv^{(j)}_{min}(\R))\, .
\label{topchsfp2}
\end{equation} 
In practice, the $j=2$ value agrees with the $j=\infty$ limit if the
geometrical definition is used \cite{n7,n22} for $Q_{latt}$.
The construction of $Q^{FP}$
goes the same way in Yang-Mills theories. \cite{n8,n29}

\subsection{There are no topological artifacts if 
FP operators are used}

We show now that using the FP action and the FP topological charge, the action
of {\it any} configuration (classical solution or not) satisfies
\begin{equation}
{\cal A}^{FP}(\R) \geq 4\,\pi |Q| \, ,
\label{ineq1}
\end{equation} 
like in the continuum classical theory. There are no 
configurations with topological charge whose action is below that of the
instanton solutions.

In the equation
${\cal A}^{FP}(\R)={\cal A}^{FP}(\Sv_{min})+T^{(\infty)}_\sigma(\R,\Sv_{min})$ we can
express ${\cal A}^{FP}(\Sv_{min})$ again in terms of the sum of the action and the
kernel on the next finer lattice, etc. Iterating further we get arbitrarily
close to the continuum, where the inequality eq.~(\ref{ineq1})
is certainly satisfied. Since the kernels are non-negative (Sect.~9.1),
eq.~(\ref{ineq1}) follows.

\section{Chiral symmetry on the lattice}

Although it has not been demonstrated rigorously, it is strongly believed that
chiral symmetry is spontaneously broken in QCD. Without this dynamical effect
QCD can not describe the observed properties of low energy hadrons. The formal
order parameter of chiral symmetry $\langle \bar\psi(x) \psi(x) \rangle$ is
equal to the trace
of the quark propagator averaged over the gauge field
configurations. The trace is proportional to the quark mass $m_q$, hence
in the $m_q \rightarrow 0$ chiral limit we can get a non-zero value only if
there is an appropriate accumulation of small eigenvalues of the massless
Dirac operator \cite{BC}. Actually, there exists a rigorous theorem \cite{AS} concerning
the zero eigenvalues of the eigenvalue equation in the continuum (we consider
one flavour, $N_f=1$)
\begin{equation}
\gamma_\mu\bigl(\partial_\mu -i\,A_\mu(x)\bigr) \psi^{(\lambda)}(x)
=\lambda \psi^{(\lambda)}(x)\,. 
\label{deigen}
\end{equation}
According to this theorem, 
if the background gauge field has a non-zero topological charge $Q$, then
there necessarily exist zero eigenvalues. The corresponding eigenvectors are
$\gamma_5$ eigenstates: $\gamma_5 \psi^{(0)}_{R,L}= \pm \psi^{(0)}_{R,L}$. If we
denote by $n_R$ and $n_L$ the number of right and left handed eigenvalues, the
theorem says
\begin{equation}
n_L-n_R=Q \, .
\label{rlq}
\end{equation} 
This result might have important consequences on the low energy properties of QCD. 
A possible intuitive picture of a typical gauge configuration in QCD is that 
of a gas or liquid of instantons and anti-instantons with quantum fluctuations
\cite{DS}. For such configurations, the Dirac operator is expected to have
a large number of quasi-zero modes which could be responsible for spontaneous
chiral symmetry-breaking.

Typical gauge field configurations contributing to the path integral are not
smooth, however. In addition, the regularization which is necessary to define
the quantum field theory breaks some other conditions of the index theorem
as well. Standard lattice formulations, for example, violate the conditions of
the index theorem in all possible ways: the topological charge of coarse gauge
configurations is not properly defined and the Dirac operator breaks 
chiral symmetry in an essential way. 
Close to the continuum some trace of the index theorem can be
identified \cite{SV,GH}. By improving the chiral behaviour of the Wilson
action a strict connection between the small real eigenvalues of the
Dirac operator and the geometric definition of the topological charge can be
established even on coarse configurations \cite{hern}.
In general, however, the expected chiral
zero modes are washed away and, even worse, the real modes which occur create
serious difficulties in simulations. They lead to 'exceptional configurations'
where the quark propagator
becomes singular when the bare mass is still distinctly different from its
critical value \cite{EX}.
 
Even in the continuum limit, the explicit chiral symmetry breaking in the
action leaves its trace behind in the quantum theory creating technical
difficulties. The quark mass receives an additive renormalization, hence the
bare quark mass should be tuned to its ($\beta$ dependent) critical value to
get the pion massless. Against the expectation that a partially conserved
current is not renormalized, the axial vector current suffers also from 
renormalization. Further,
there is mixing between operators in nominally different chiral
representations creating a special difficulty in weak matrixelement
calculations.

Fixed point actions offer a bold solution to all these problems. The good
chiral behaviour of the FP action follows essentially from the fact that its
fermion part satisfies the Ginsparg-Wilson remnant chiral symmetry condition
\cite{GW}, for which no solution was known previously. This is the point,
where the FP action meets a seemingly unrelated approach to
overcome the problems of lattice regularization concerning chiral
symmetry \cite{Kap,Neu}. The purpose of this section is to discuss these recent
developments.

\subsection{'On-shell' symmetries}

As the example of instanton solutions illustrates, scale invariance, which is
obviously broken by the lattice, remains a symmetry on the classical solutions
of the FP action. This observation can be generalized \cite{onsh}. For any
symmetry of the classical countinuum theory (arbitrary translation, rotation,
chiral rotation, etc.) a representation can be defined on the classical
solutions.

For the arguments, let us use the language of the non-linear
$\sigma$-model. Assume that the confuguration $\R$, which lives on a lattice
with lattice unit $a_0$,
is a solution of the
Euler-Lagrange equations of motion of the FP action. As we have discussed in
Sect.~9.1, the minimizing configuration $\Sv^{(1)}_{min}(\R)$ associated with $\R$ is
also a solution. This configuration is defined on a lattice with unit $a_1=a_0/2$.
We can iterate this process: to the solution $\Sv^{(1)}_{min}(\R)$
there corresponds again a minimizing configuration $\Sv^{(2)}_{min}(\Sv^{(1)}_{min}(\R))$
which lives on a lattice
with lattice unit $a_2=a_0/2^2$, etc. Iterating a large number of times we get
arbitrarily close to the continuum
\begin{equation}
\lim_{j \to \infty} \Sv^{(j)}_{min} = \Sv(x,\R) \, ,
\end{equation}
where $\Sv(x,\R)$ is a solution in the continuum. On this solution all the
symmetry transformations of the continuum theory are well defined. Take, for
example an infinitesimal translation: $\Sv \to \Sv^\epsilon$ where
$\Sv^\epsilon(x,\R)=\Sv(x+\epsilon,\R)$. The configuration $\Sv^\epsilon$ is
also a solution which we can block back to the starting lattice with unit
$a_0$ producing a a new solution $\R^\epsilon$ there \footnote{As discussed in
Sect.~9.1 blocking a solution might not lead to a solution again if we land on
a very coarse lattice. Assume, the original solution $\R$ is an instanton with
radius $\rho$ slightly larger than the critical size $\sim 0.7 a$. Blocking
back the corresponding continuum solution after translation, we get a
translated instanton with the same $\rho$ on the coarse lattice. Since the
critical size depends slightly on the position of the instanton on the lattice
\cite{n7,n22}, the translated small instanton might 'fall through the lattice'. To
avoid this complication, assume that the solution $\R$ is away from this
critical region.}. The process
\begin{equation}
\R_n \to \Sv(x,\R) \to \Sv(x+\epsilon,\R) \to \R^{\epsilon}_n \, ,
\end{equation}
defines a representation of the group of continuous translations. It goes
similarly for other symmetry transformations of the classical continuum
theory.

We emphasize that the procedure above makes sense only for configurations which are
solutions of the classical equations of motion. Therefore the name: 'on-shell'
symmetry.

\subsection{The index theorem on the lattice (I)}

Consider the eigenvalue equation of the FP Dirac operator
\begin{equation}
\sum_{n'} h^{FP}(U)_{n,n'} \psi^{(\lambda)}_{n'}
=\lambda \psi^{(\lambda)}_n \, .
\label{deigen1}
\end{equation} 
The index theorem refers to the $\lambda=0$ eigenvalues, in which case the
eigenvalue equation above is reduced to classical Euler-Lagrange equation
(Dirac equation). Due to the 'on-shell' symmetries discussed in Sect.~10.1, we
suspect that the FP action will lead to an exact index theorem on the lattice
independently whether the background gauge configuration is smooth, or
coarse. This is really so \cite{unp}. The proof is based on the FP equations 
for the gauge action (eq.~(\ref{fpym})) and for the Dirac operator (eq.~(\ref{fph1}).

Consider a gauge configuration $V$ on a lattice with lattice unit
$a_0$. Assume that $V$ has a topological charge $Q$. As discussed in
Sect.~9.3, the FP topological charge is defined through a sequence of minimizing
configurations $U^{(1)}_{min}(V),
U^{(2)}_{min}(U^{(1)}_{min}(V)),\ldots$. Here $U^{(j)}_{min}$ lives on a
lattice with lattice unit $a_j=a_0/2^j$. By construction all these
configurations have the same topological charge $Q$, which is the charge of
$U^{(\infty)}_{min}$ in the continuum, at the end of the sequence.

Assume that the FP Dirac operator $h^{FP}$ has a zero eigenvalue for the
background field $V$
\begin{equation}
\sum_{n'_B} h^{FP}(V)_{n_B,n'_B} \chi_{n'_B} = 0 \, .
\label{diracv}
\end{equation}  
Eq.~(\ref{diracv}) is the Dirac equation, i.e. the classical Euler-Lagrange
equation of motion. We proceed now as we did in the case of the classical 
solutions in Sect.~9.1. Taking the variation of eq.~(\ref{fph1}) 
with respect to $\bar{\chi}_{n_B}$ and using
eqs.~(\ref{hdef},\ref{diracv},\ref{tfqcd}) we find that $\psi^{st}_n$, which
makes the r.h.s. of eq.~(\ref{fph1}) stationary, satisfies the Dirac equation
for the background field $U^{(1)}_{min}(V)$. Since eq.~(\ref{fph1}) is
quadratic in the fermion field, the stationary configuration is unique. Hence
the opposite statement is also true: if $\psi_n$ is a solution of the Dirac
equation for the background field $U^{(1)}_{min}(V)$ then
$\chi_{n_B}=b_f \sum_{n}\omega(U)_{n_B,n} \psi_n$ is a solution on the coarse
lattice for the configuration $V$.

This result can be iterated towards finer lattices. Having a zero mode
eigenfunction $\psi^{st}_n$ for $U^{(1)}_{min}$ there is a zero mode on the
next finer lattice for the configuration  $U^{(2)}_{min}$, etc. At every step
we can also move in the opposite direction. Going up, we finally get to a very
smooth configuration for which the Atiyah-Singer theorem is applicable. This way
we get that for each gauge configuration of the sequence, the FP Dirac
eigenvalue problem has the same number of zero modes and with the same
helicities as in the continuum. The statement on the helicities follows from
the relation between the eigenvectors of the $i$-th and $(i-1)$-th step
\begin{equation}
\psi^{st}_{n_{(i-1)}} = b_f\, \sum_{n_{(i)}}
\omega(U^{(i)}_{min})_{n_{(i-1)},n_{(i)}}\psi^{st}_{n_{(i)}} \, ,
\end{equation} 
and from the fact that $\omega$ is trivial in Dirac space (Sect.~5.3).

We have to emphasize that for the validity of the index theorem we need a
classically perfect regularization both in the fermion and in the gauge sector.
This is also obvious from the fact that the index theorem connects a quantity
defined in the gauge sector (topological charge $Q$ of the gauge
configuration) with that coming from the fermion sector (the number of R and L
zero eigenmodes).

\subsection{Free FP fermions}

A good starting point to understand the chiral properties of the FP action is
the case of free FP fermions. The FP of free theories (scalar, gauge, or
fermions) can be constructed analytically and their properties can be studied 
in every detail \cite{bell,Bal,hn,uwe1}, Sect.~8.

Write $U_{min}=1$ in the recursion relation eq.~(\ref{transfh}) and consider
the example of the simple block transformation $\omega(2n_B-n)=2^{-d}$
if $n$ is in the hypercube whose center is indexed by $n_B$ and zero
otherwise, Sect.~5.3 . Starting with the Wilson action (with $m_q=0$)
on a very fine lattice, the iterated RGT drives the system to a FP. 
In Fourier space the FP propagator reads \cite{GW,uwe1} 
\begin{equation}
\label{freefp}
h^{-1}(q)= \sum_{l \in Z^d} \frac{\gamma_\mu (q_\mu + 2\pi l_\mu)}
{(q+2\pi l)^2} \prod_{\nu}\frac{\sin^2({q_\nu \over 2})}
{({q_\nu \over 2}+\pi l_\nu)^2} + {{2} \over {\kappa_f}} \, .
\end{equation}
Here $q$ is defined over the Brillouin zone.
The poles of this propagator give the exact continuum spectrum, as
expected. There is no doubling. (The massive doublers run out to infinity
under the repeated RGT.) The action is a periodic analytic function of $q$, in
configuration space it is local. Complying with the no-go theorem \cite{nn},
the propagator, and so the action, break chiral symmetry. This symmetry
breaking is, however, very special:
\begin{equation}
\{ h^{-1},\gamma^5 \}=\frac{4}{\kappa_f}\gamma^5 \,.
\label{chirfree}
\end{equation}
The chiral symmetry breaking term
$2/\kappa_f$ in the propagator comes entirely from the block
transformation. The effect of the chiral symmetry breaking in the propagator
shows up only at $n=0$ in configuration space. No physical properties are
influenced by that \footnote{At $\kappa_f=\infty$ the FP action becomes chiral
symmetric, but at the same time, non-local\cite{uwe1} in consistency with the
no-go theorem \cite{nn}}.

\subsection{The Ginsparg-Wilson remnant chiral symmetry condition}

More than fifteen years ago, Ginsparg and Wilson \cite{GW} suggested to elevate the free
field result eq.~(\ref{chirfree}) into a general principle which defines
(remnant) chiral symmetry on the lattice for interacting theories also. The
condition is derived in Ref.~\cite{GW} by considering RG transformations 
starting out of the
continuum theory. Since, however, in the derivation the assumption is made that the
blocked action is quadratic in the fermion fields (which is true only in a
free field theory) it might be better to consider the GW remnant chiral symmetry
condition as an genial, intuitive suggestion: the lattice regularization is
expected to preserve the consequences of chiral symmetry in interacting
theories also (in particular in QCD) if the Dirac operator satisfies
\begin{equation}
\label{1}
\frac {1}{2}\{ h^{-1}_{n n'} (U), \gamma^5 \} = 
R_{n n'}(U) \gamma^5 \,,
\end{equation}
or equivalently
\begin{equation}
\label{2}
\frac {1}{2}\{ h_{n n'}, \gamma^5 \} = 
(h \gamma^5 R h)_{n n'} \,.
\end{equation} 
Here $h(U)^{-1}$ is the quark propagator over the background gauge field $U$
and the matrix $R_{nn'}$ is trivial in Dirac space. Both $h(U)_{nn'}$ and
$R(U)_{nn'}$ are local. 
The locality of $R$ in the remnant chiral symmetry
condition is absolutely essential. It is also highly non-trivial, since
$h^{-1}$ is a non-local quantity.

Ginsparg and Wilson have shown\cite{GW} that eq.~(\ref{1}) implies the
correct triangle anomaly on the lattice. In the presence of gauge fields,
however, no solution was found by the authors and the paper remained
practically unnoticed for a long time.

\subsection{Solutions of the remnant chiral symmetry condition}

It has been noticed recently that the FP action of QCD satisfies the remnant
chiral symmetry condition eq.~(\ref{1}) \cite{FPGW,HLN}. This result is a
direct consequence of the FP equations eqs.~(\ref{fpym},\ref{transfh}).

Eq.~(\ref{transfh}) can be solved by iteration. 
If the scale change $s$ of the
transformation is (infinitely) large, the fixed point can be reached in a
single renormalization group step. In this case, the fine field $U_{min}$ 
lives on
an infinitely fine lattice, and it is smooth. The Dirac operator
$h$ over this field is arbitrarily close to the massless continuum Dirac
operator. Since $\omega$ is trivial in Dirac space, the second term on the
r.h.s. of eq.~(\ref{transfh}) is chiral invariant, which leads to the remnant
chiral symmetry condition in eq.~(\ref{1}) with 
$R_{n n'}=1/\kappa_{\rm f}\cdot\delta_{n n'}$. 
If $s$ is finite, one can start again on an infinitely fine lattice, 
but the transformation should be iterated. One obtains in every iteration 
step some contribution to the chiral symmetry breaking part of $h^{-1}$, 
and a non-trivial, but local $R$ builds up in this case.

Very recently, another solution of eq.~(\ref{1}) was presented
\cite{NeuGW}. This solution grew out the overlap formalism \cite{Neu,Neu3} and
seems to be unrelated to renormalization group considerations. The solution
has the structure $h(U)_{n,n'}=\delta_{n,n'}+V(U)_{n,n'}$ (only the space
indices are indicated explicitly), 
where the unitary matrix $V$ has the form
\begin{equation}
\label{neu}
V=X \frac{1}{\sqrt(X^\dagger X)}\, .
\end{equation}
Here $X(U)_{n,n'}$, $X^\dagger=\gamma^5 X \gamma^5$ is an ultralocal interaction,
whose form is explicitly given \cite{NeuGW}. In spite of the square-root in the 
definition of $V$, the action is obviously local in the free case ($U=1$), and
is, presumably, also local in general. It is a little exercise to show that
the Dirac operator $h$ defined this way satisfies the remnant chiral symmetry
condition eq.~(\ref{1}) with $R_{n,n'}=1/2\delta_{n,n'}$. An advantage of
this solution is that it is explicitly given, while to find the FP action the
defining classical equations have to be solved. A disadvantage is that it is
significantly less shortranged than the optimized FP Dirac operators. They
decay exponentially with a constant $\gamma=O(3)$ in the exponent, while the
solution discussed above decays only with $\gamma=O(1)$. Another difference
is that the overlap formalism addresses the issue of chiral symmetry only,
while the FP action is classically perfect. For free fermions, for example,
the action defined by eq.~(\ref{freefp}) has an exact spectrum, while the
spectrum of the Dirac operator above is similar to that of the Wilson action.
Nevertheless, it is interesting that the structure of eq.~(\ref{neu}) alone assures
the GW condition which might be a very useful observation in itself.

\subsection{Currents}

In the following sections we shall need different currents which we construct
here and study some of their properties. Consider a fermion action which is
quadratic in the fermion fields
\begin{equation}
{\cal A}_f(\bar{\psi},\psi,U) = \sum_{m',n'}\bar{\psi}_{m'} h_{m',n'}(U)\psi_{n'}\, .
\label{cur1}
\end{equation}
Most of our discussion will be general, in particular we do not assume that
${\cal A}_f$ is a FP action, or that it satisfies the Ginsparg-Wilson relation.

\noindent A) U(1) gauge field, vector current, sum rules

We shall use the example of the electromagnetic current to illustrate the
procedure. Consider a $U(1)$ gauge field (electromagnetic field) coupled to a
charged fermion and construct the current which belongs to the global vector
symmetry. Following the usual steps of defining a Noether current we consider
the local transformation
\begin{equation}
\delta \psi_n = i \epsilon_n \psi_n \,, \qquad
\delta \bar{\psi}_n = - \bar{\psi}_n i \epsilon_n \, ,
\label{cur2}
\end{equation}
and obtain
\begin{equation}
\delta {\cal A}_f = -i \sum_n \epsilon_n \bar{\nabla}_\mu J_\mu (n) \, ,
\label{cur3}
\end{equation}
where
\begin{equation}
\label{cur4}
\bar{\nabla}_\mu J_\mu (n)= \sum_{m'} [ 
\bar{\psi}_n h(U)_{n,m'} \psi_{m'} - 
\bar{\psi}_{m'} h(U)_{m',n} \psi_n ] \,. 
\end{equation}
The derivative $\bar{\nabla}_\mu$ is defined as
\begin{equation}
\label{cur5}
{\bar \nabla}_\mu f(n)=f(n)-f(n-\hat\mu) \, .
\end{equation}
We want to find a vector field $J_\mu (n)$ which satisfies
eq.~(\ref{cur4}). A solution is
\begin{equation}
\label{cur6}
J_\mu(n) = 
 - i \sum_{m',n'} \bar{\psi}_{m'}
 \frac{\delta}{\delta A_\mu(n)} 
h(U)_{m',n'} \psi_{n'}  \,,
\end{equation}
where $A_\mu(n)$ is the vector potential, $U_\mu(n)=e^{i A_\mu(n)}$. The
action ${\cal A}_f$ is gauge invariant: $\bar{\psi}_{m'}$ and $\psi_{n'}$ are connected
by a product of $U$ fields along paths running between the fermion
offset $m',n'$. Consider a fix fermion offset $m',n'$. Since
\begin{equation}
\label{cur7}
\frac{\delta}{\delta A_\mu(n)}U_\nu(m)= i \delta_{\mu,\nu} \delta_{nm} U_\mu(n) 
\end{equation}
we get that every path (between $m',n'$) going through the link
$(n,n+\hat{\mu})$ gives a contribution to $J_\mu (n)$ which is equal to the
contribution of this path to $h_{m',n'}$ itself, times $(s_+-s_-)$,
where $s_+ (s_-)$ is the number this path runs through $(n,n+\hat{\mu})$ in
positive (negative) direction. This $U(1)$ current is obviously gauge
invariant.

We want to show that the current in eq.~(\ref{cur6}) satisfies
eq.~(\ref{cur4}). Consider $\bar{\nabla}_\mu J_\mu (n)$
\begin{equation}
\label{cur8}
\bar{\nabla}_\mu J_\mu (n)=(J_1(n)-J_1(n-\hat{1})) + (J_2(n)-J_2(n-\hat{2}))+
\dots  \,.
\end{equation}
To $J_1(n-\hat{1})$ all the paths contribute which go through the link
$(n-\hat{1},n)$. Take first $n \ne m', n \ne n'$. All the paths which run into the
point through the link $(n-\hat{1},n)$, (and so contributing to
$J_1(n-\hat{1})$), will run out of this point through one of the links
$(n,n+\hat{\nu})$ (and so contributing to $J_\nu(n)$), etc. These
contributions cancel. A similar consideration shows that if $n=m'$, or $n=n'$
the contribution to $\bar{\nabla}_\mu J_\mu (n)$ is $h_{n,m'}$ and
$-h_{m',n}$, respectively, leading to eq.~(\ref{cur4}). 

If the action ${\cal A}_f$ is invariant under the reflection of one of the
coordinate axis, then $J_\mu(n)$ in eq.~(\ref{cur6}) will transform like the
vector potential $A_\mu(n)$ under this transformation, i.e. like a lattice
vector field, as it should.

Consider the current with $U=1$. The current is translation invariant in this
case and we can write
\begin{equation}
\label{cur9}
J_\mu (n)= \sum_{m',n'} \bar{\psi}_{m'}
\Gamma_\mu(m'-n,n'-n) \psi_{n'}  \,.
\end{equation}
Although the form of the current depends on the details of the Dirac operator,
the following sum rules, which test the long-distance properties of the current, 
are generally valid if the action preserves cubic symmetry:
\begin{equation}
\label{cur10}
\sum_m \Gamma_\mu(m,m-l)=-l_\mu h_l \,,
\end{equation}
\begin{equation}
\label{cur11}
\sum_m m_\nu \Gamma_\mu(m,m-l)=- \frac{l_\mu(l_\nu+\delta_{\mu\nu})}{2} h_l\,,
\end{equation}
where we introduced the notation $h(U=1)_{m,n}=h_{m-n}$. Since these sum rules
enter the derivation of the axial anomaly \cite{GW} \footnote{In
Ref.~\cite{GW} the current was constructed in a particular way, and the sum
rules above were shown explicitly to be satisfied. The particular form of the
current in Ref.~\cite{GW} is not convenient in different theoretical
considerations,however.}, we sketch their derivation.

Fix the fermion offset $m'-n'=-l$. Using translation symmetry, we can take
$m'=0,\,n'=l$. Summing over $n$ 
\begin{equation}
\label{cur12}
\sum_n \Gamma_\mu(-n,l-n) \,,
\end{equation}
we have to consider all the paths in $h_{-l}$ and for each path count the links
going in the $\pm \hat{\mu}$ direction with the weight $\pm 1$. Since the paths
run from $(0,0,\dots,0)$ to $(l_1,l_2,\dots,l_d)$, each path gives $l_\mu$
leading to 
\begin{equation}
\label{cur13}
\sum_n \Gamma_\mu(-n,l-n)=l_\mu h_{-l}
\end{equation}
Taking $l \to -l$ and changing the summation variable $-n \to m$ we get 
the first sum rule in eq.~(\ref{cur10}).

Consider now
\begin{equation}
\label{cur14}
\sum_n n_\nu \Gamma_\mu(-n,l-n) \,.
\end{equation}
Due to hypercubic symmetry, a path and its reflection to the point in the
middle of $\bar{l}$ contribute the same amount to $h_{-l}$. 
Take the contribution
of an arbitrary path and of its central reflected path to eq.~(\ref{cur14}).
Consider first $\mu \neq \nu$. A path going through $(n,n+\hat{\mu})$
contributes $n_\nu$. The reflected path gives
$l_\nu-n_\nu$, so the average is $l_\nu/2$. Summing over all the links in the
$\hat{\mu}$ direction gives $l_\mu l_\nu/2$. Consider now $\mu= \nu$. The link
$(n,n+\hat{\mu})$ on the path contributes $\pm n_\mu$ depending on the
direction of running through. Summing over all the links gives
$l_\mu(l_\mu-1)/2$. We obtain
\begin{equation}
\label{cur15}
\sum_n n_\nu \Gamma_\mu(-n,l-n)=\frac{l_\mu(l_\nu-\delta_{\mu\nu}}{2}h_{-l} \,.
\end{equation}
Taking $l \to -l$ and changing the summation variable $-n \to m$ we get 
the second sum rule in eq.~(\ref{cur11}).

\noindent B) QCD flavour currents

Consider eq.~(\ref{cur1}) with $N_f$ flavours and assume that $h$ is flavour
symmetric. Denote the flavour generators by $\tau^a,\,a=1,\dots,N_f^2-1$ with
\begin{equation}
\label{cur16}
[\tau^a,\tau^b]=i f^{abc}\tau^c, \qquad 
tr(\tau^a\tau^b)=\frac{1}{2}\delta_{ab}\,.
\end{equation}
The fermion action has a $U(N_f)$ flavour vector symmetry. As we discussed
before, the axial symmetry is broken, even if we set the bare mass to zero. 
As usual, the axial currents are constructed from the chiral symmetric part of
the action using 
${h}_{\rm SYM}=\frac{1}{2} [{h},\gamma^5] \gamma^5$. 
There is a certain freedom in constructing the
vector currents. Using the full action, conserved vector currents are
obtained. On the other hand, the asymmetry between the vector and axial case
generates strange terms in current algebra relations and it requires extra
work to show that they go away. Let us construct
both currents with ${h}_{\rm SYM}$. 
Using the steps we discussed in A) we get
\begin{equation}
\label{cur17}
\bar{\nabla}_\mu V_\mu^a (n)= 
\bar{\psi}_n \tau^a \left( {h}_{\rm SYM} \psi \right)_n - 
\left( \bar{\psi} {h}_{\rm SYM} \right)_n \tau^a \psi_n \,, 
\end{equation} 
\begin{equation}
\label{cur18}
\bar{\nabla}_\mu A_\mu^a (n)= 
- \bar{\psi}_n \tau^a \left( \gamma^5 {h}_{\rm SYM} \psi \right)_n - 
\left( \bar{\psi} {h}_{\rm SYM}\gamma^5 \right)_n \tau^a \psi_n \,. 
\end{equation}
None of these currents are conserved if the equations of motion
${h} \psi =0$ is used.

In order to find the currents themselves we introduce the flavour gauge
matrices $W_\mu(n)= 1 + i w_\mu^a(n)\tau^a +\dots$ and extend the
product of colour $U$ matrices in eq.~(\ref{cur1}) along the paths between the
fermion offsets $m',n'$ by the product of $W$ matrices along the same
paths. Then the vector current is defined by
\begin{equation}
\label{cur19}
V_\mu^a(n) = -i
\sum_{m',n'} \bar{\psi}_{m'} 
\left. \frac{\delta}{\delta w_\mu^a(n)} 
({h}_{\rm SYM})_{m' n'}(U,W) \psi_{n'} \right|_{w=0} \,.
\end{equation} 
This current satisfies eq.~(\ref{cur17}) as can be shown using the same
arguments we used in A). 
Eq.~(\ref{cur19}) can also be written as 
\begin{equation}
\label{cur20}
V_\mu^a(n) =
\sum_{m' n'} \bar{\psi}_{m'} \tau^a \Gamma_\mu(m',n';n;U)
\psi_{n'} \, ,
\end{equation} 
where $\Gamma_\mu$ is trivial in flavour space. Similar considerations lead to
the axial current 
\begin{equation}
\label{cur21}
A_\mu^a(n) =
\sum_{m' n'} \bar{\psi}_{m'} \tau^a \Gamma_\mu^5(m',n';n;U)
\psi_{n'} \, ,
\end{equation} 
where 
\begin{equation}
\label{10}
\Gamma_\mu^5 = \Gamma_\mu \gamma^5 = - \gamma^5 \Gamma_\mu \, .
\end{equation}
The flavour singlet vector and axial currents are obtained from
eqs.~(\ref{cur17},\ref{cur18}) by replacing $\tau^a$ by the unit matrix in
flavour space. 

\subsection{The index theorem on the lattice (II)}

In Sect.~10.2 we demonstrated that the index theorem is valid on the lattice if
the FP action is used. The method was to connect the lattice problem with that
in the continuum (where the index theorem is true) using the FP equations.
One can prove, however, the index theorem directly on the lattice without
refering to results obtained earlier in the continuum \cite{HLN}. 
We shall discuss this
derivation briefly since it leads to additional results and insight.

We shall suppress the index $FP$ on the fixed point Dirac operator $h^{FP}$ in
the following.
To avoid some singularities at intermediate steps of the calculation
we introduce a small quark mass which will be sent to zero at the end:
\begin{equation}
\label{in1}
h_{m' n'}(U) \to \hat{h}_{m' n'}(U)=h_{m' n'}(U) + 
m_{\rm q}\, \delta_{m' n'} \,.
\end{equation}
Due to eq.~(\ref{2}) the chiral symmetry breaking part of $\hat{h}$ satisfies
the relation
\begin{equation}
\label{in2}
\frac{1}{2} \{ \hat{h}_{m' n'}, \gamma_5 \} \gamma_5 = 
 (h^\dagger R h)_{m' n'} +  m_q \, \delta_{m' n'} \,.
\end{equation}
This equation plays a basic role in deriving the results.

Consider the flavour singlet axial vector current (take $N_f=1$) and write
\begin{equation}
\label{in3}
h_{\rm SYM}=\frac{1}{2}[h,\gamma_5]\gamma_5=
\hat{h}-\frac{1}{2}\{\hat{h},\gamma_5 \} \gamma_5 \,.
\end{equation}
Using eq.~(\ref{in2}) we obtain
\begin{eqnarray}
\label{in4}
 \bar{\nabla}_\mu A_\mu (n)  & = &
 \bar{\psi}_n \gamma_5 ( \hat{h}\psi )_n + 
( \bar{\psi} \hat{h})_n \gamma_5 \psi_n  \nonumber \\
 & & -  \bar{\psi}_n \gamma_5 \left( h^\dagger R h \psi \right)_n + 
\left( \bar{\psi} h^\dagger R h\right)_n \gamma_5\psi_n  \\
 & & -2 m_{\rm q}  \bar{\psi}_n \gamma_5 \psi_n  \,.\nonumber
\end{eqnarray}
The first term on the r.h.s. is proportional to the equation of motion 
$\hat{h}\psi = 0$, the second comes from the remnant chiral symmetry 
condition, while the last term is due to the explicit symmetry breaking.
Consider the expectation value of this equation over a background gauge
field. For any operator ${\cal O}$ this expectation value is defined as
\begin{equation}
\label{in5}
\langle {\cal O}(\bar{\psi},\psi,U) \rangle 
= \frac{1}{Z(U)} \int D\bar{\psi} D\psi {\cal O}(\bar{\psi},\psi,U)
\exp\{ -\sum_{m,n} \bar{\psi}_{m} \hat{h}_{m n}(U)\psi_{n} \} \,,
\end{equation}
where $Z(U)$ is given by the analogous integral without the
${\cal O}(\bar{\psi},\psi,U)$ factor. The contribution from the first term
is zero.
Indeed,
\begin{equation}
\label{in6}
\langle \bar{\psi}_n \gamma_5 ( \hat{h}\psi )_n \rangle =
-\frac{1}{Z(U)} \int D\bar{\psi} D\psi \, \bar{\psi}_n \gamma_5 
\frac{\delta}{\delta\bar{\psi}_n}
\exp\{ -\sum_{m',n'}\bar{\psi}_{m'} \hat{h}_{m' n'} \psi_{n'}\}\,,
\end{equation}
and partial integration by $\bar \psi$ gives then ${\rm tr}\gamma_5=0$. 
Integrating out the
fermions in the expectation value of the third term in eq.~(\ref{in4}) gives 
$2m_q {\rm tr}(\gamma_5{\hat h}^{-1}_{n n})$, where the trace is over colour
and Dirac space.
We insert here 
the closure relation of eigenvectors of the hermitian matrix $H=h\gamma_5$. 
It is easy to show using the Ginspar-Wilson relation that the zero eigenvalue 
eigenfunctions of $H$ (if there are any) are chiral and they are 
zero eigenvalue eigenfunctions  of $h$ as well. 
In the $m_q \rightarrow 0$ limit only these eigenfunctions
contribute leading to
\begin{equation}
\label{in7}
\lim_{m_{\rm q}\to 0} m_{\rm q}
\langle \bar{\psi}_n \gamma_5 \psi_n \rangle =
\sum_{j=1}^{n_{\rm L}} \psi_{\rm L}^{(j)*}(n) \psi_{\rm L}^{(j)}(n)
-\sum_{i=1}^{n_{\rm R}} \psi_{\rm R}^{(i)*}(n) \psi_{\rm R}^{(i)}(n) \,,
\end{equation} 
where $\psi^{(i)}_R(n)$, $\psi^{(j)}_L(n)$ are the normalized 
$\lambda=0$ right- and left-handed eigenfunctions of the eigenvalue problem 
of the Dirac operator. Finally, integrating out the fermions in the 
expectation value of the second
term in eq.~(\ref{in4}) and summing over n we obtain
\begin{equation}
\label{in8}
0= 2\, {\rm Tr}(\gamma_5 h R) + 2(n_R - n_L)  \, ,
\end{equation} 
where the trace ${\rm Tr}$ is over colour, Dirac and configuration space.
The first term is a pseudoscalar gauge invariant functional of 
the background gauge field. For smooth fields $U$ its density
is expected to be proportional to $\epsilon^{\mu\nu\alpha\beta}
F_{\mu\nu}^{a}(n)F_{\alpha\beta}^{a}(n)$. Really, one can show \cite{HLN}
\begin{equation}
\label{in9}
- \langle \bar{\psi}_n \gamma_5 \left(h^\dagger R h \psi \right)_n + 
\left( \bar{\psi} h^\dagger R h\right)_n \gamma_5 \psi_n \rangle \to
\frac{1}{32\pi^2} \epsilon^{\mu\nu\alpha\beta}
F_{\mu\nu}^{a}(n)F_{\alpha\beta}^{a}(n) 
\end{equation}
up to terms which are of higher order in the vector potential and/or 
derivatives. This result follows from Ref.~\cite{GW} where the
anomaly of the flavour singlet axial current was derived
using the remnant chiral symmetry
condition and the sum rules on the currents we discussed in Sect.~10.6.
Further, it can be shown that the first term in eq.~(\ref{in8}) is a FP
operator: it is reproduced under a RGT. It follows than that
\begin{equation}
\label{in10}
{\rm Tr}(\gamma_5 h R) =Q^{\rm FP} \, ,
\end{equation}
where $Q^{\rm FP}$ is the FP topological charge defined in
the Yang-Mills theory, Sect.~9.3. Eq.~(\ref{in10}) is a remarkable connection
between the gauge and fermion parts of the FP theory. 

\subsection{The spectrum of the Dirac operator satisfying the remnant chiral
  symmetry condition}

\begin{figure}[htb]
\begin{center}
\leavevmode
\epsfxsize=110mm
\epsfbox{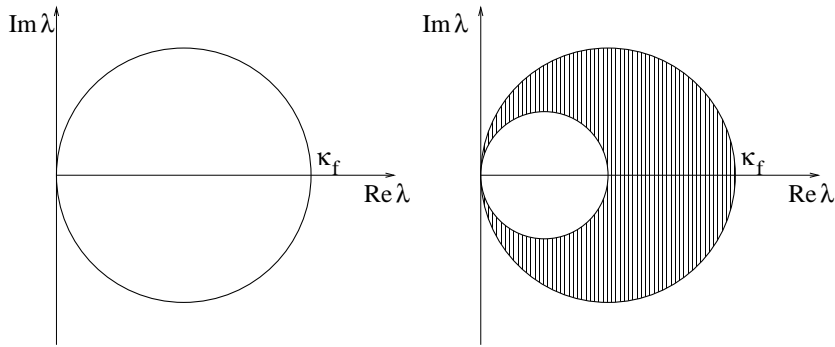}
\end{center}
\caption{The spectrum of the fixed point Dirac operator:
a) for the special case $R_{nn'}=1/\kappa_{\rm f}\cdot\delta_{nn'}$
the eigenvalues lie on a circle,
b) for a more general set of block transformations the eigenvalues
lie between the two circles.}
\label{fig5}
\end{figure}

Let $\psi$ be a normalized solution of the eigenvalue equation,
$h\psi=\lambda\psi$. Eq.~(\ref{2}) (written in the form
$\frac{1}{2}(h+h^\dagger) = h^\dagger Rh$) implies 
${\rm Re}\lambda=|\lambda|^2(\psi,R\psi)$. 
For the special case $R_{n n'}=1/\kappa_{\rm f}\cdot\delta_{n n'}$
the eigenvalues $\lambda$ lie on a circle of radius 
$\kappa_{\rm f}/2$ touching the imaginary axis, as shown in fig.~\ref{fig5}.
In general, the eigenvalues lie between two circles tangent to the imaginary
axis - the circle described previously and a smaller one as shown in
fig.~\ref{fig5}. The smaller circle has a finite radius which follows from the
locality of $R$. 
This property excludes exceptional configurations, appearing e.g. for
the Wilson action and causing serious problems in numerical simulations.
As a consequence of $h^\dagger=h \gamma_5 h$ the solutions with 
${\rm Im}\lambda\ne 0$ come in complex conjugate pairs and satisfy
$(\psi,\gamma_5\psi)=0$. Due to eqs.~(\ref{1},\ref{2}), those with 
$\lambda=0$ are chiral eigenstates, $\gamma_5\psi=\pm\psi$.

\subsection{Lattice QCD without tuning, mixing and current renormalization}

As we discussed at length before,
lattice regularized, local QCD actions break chiral symmetry under 
general conditions. Wilson fermions have 
a dimension five
symmetry breaking operator whose effect on the physical predictions goes to
zero in the continuum limit. It leaves, however its trace behind in the form
of additive quark mass renormalization, axial current renormalization and
mixing between operators in nominally different chiral representations. There
is a significant recent progress in calculating these renormalizations in a
theoretically controlled non-perturbative way \cite{NONP}. 
Nevertheless, the situation is
not really pleasing theoretically and the technical difficulties are
also significant.

On the other hand, if the Dirac operator satisfies the remnant chiral symmetry
condition, then all the physical consequences of chiral symmetry survive the
lattice regularization without quark mass tuning, without the mixing 
of operators in different chiral representation and without flavour non-singlet
axial current renormalization \cite{chiral}. We do not discuss here the details
of the arguments (which are based on chiral Ward identities following the
seminal paper Ref.~\cite{CCH} but mention the intuitive reason
only. Consider the divergence of the flavour non-singlet axial vector current
for $m_q=0$
\begin{eqnarray}
\label{chi1}
 \bar{\nabla}_\mu A^a_\mu (n)  & = &
 \bar{\psi}_n \tau^a \gamma_5 ( {h}\psi )_n + 
( \bar{\psi} {h})_n \gamma_5 \tau^a \psi_n  \nonumber \\
 & & -  \bar{\psi}_n \tau^a  \left(
 h \gamma_5 R h \psi \right)_n + 
\left( \bar{\psi} h \gamma_5 R h\right)_n \tau^a \psi_n \,. 
\end{eqnarray}
The first term on the l.h.s. is proportional to the equations of motion.
Observe the special structure of the chiral symmetry breaking second term: it
contains two $h$ factors. If the divergence of the axial vector current, and
so this term enters a Green's function these two $h$ factors will cancel the
two propagators which arrive when the $\bar{\psi}_n$ and $\psi_n$ fields in 
eq.~(\ref{chi1}) are paired with the fermion fields of the other operators in
the Green's function. This breaking term, therefore is expected to produce
contact terms only which will not influence the physical predictions. 

\subsection{Exact chiral symmetry on the lattice}

Looking at the results of the previous sections, one might suspect that there
exists an underlying hidden exact symmetry on the lattice which replaces the
standard chiral symmetry in the formal continuum formulation. This is really
so \cite{lnew}. L\"uscher observed that the remnant chiral symmetry condition 
eq.~(\ref{2}) implies an exact symmetry of the fermion action, which may be
regarded as a lattice form of an infinitesimal chiral rotation. Really, the
transformation 
\begin{eqnarray}
\delta \psi_n &=& i \epsilon^a \tau^a \sum_{n'} \gamma_5
(1-Rh)_{n,n'}\psi_{n'} \,, \nonumber \\
\delta \bar{\psi}_n &=& i \sum_{n'} \bar{\psi}_{n'}
(1-hR)_{n',n} \gamma_5 \epsilon^a \tau^a \,,
\end{eqnarray}
which may be considered as a locally smeared version of the standard
$\gamma_5$ transformation, is a symmetry of the action eq.~(\ref{cur1}). There
is no contradiction with the no-go theorem \cite{nn}, since the symmetry is
realized differently as has been assumed when proving the theorem. A large
circle is closed by this observation.

One might hope that the recent developments discussed in the last sections
above
will help to solve further longstanding difficulties of lattice regularization.

\section*{Acknowledgements}
The author is indebted to the organizers of the workshops at Cambridge,
Pe\~niscola and Kyoto for the kind hospitality.


\begin{thebibliography}{99}
\bibitem{FPGW} the brief arguments are given in Ref.~\cite{has}.
\end{thebibliography}
\end{document}